\documentclass[11pt,a4paper]{article} 
\usepackage{jheppub}
\usepackage{graphicx}
\usepackage[numbers,sort&compress]{natbib}
\usepackage{amsmath}
\usepackage{amssymb}
\usepackage{relsize}

\title{Holographic meson decays via worldsheet instantons}

\author{Kasper Peeters,}
\author{Maciej Matuszewski}
\author{and Marija Zamaklar}
\affiliation{Department of Mathematical Sciences,\\
Durham University,\\
South Road,\\
Durham DH1 3LE,\\ United Kingdom.\\[.5ex]}

\emailAdd{~\\kasper.peeters@durham.ac.uk}
\emailAdd{~\\m.t.matuszewski@durham.ac.uk}
\emailAdd{~\\marija.zamaklar@durham.ac.uk}

\preprint{DCPT-18/07}

\abstract{We study meson decays using instanton methods in two string
  models. The first model is the old string model in flat space which
  combines strings and massive particles and the second is the
  holographic, Sakai-Sugimoto model.  Using the the old string model,
  we reproduce the QCD formula for the probability of splitting of the
  QCD flux tube derived by Casher-Neuberger-Nussinov (CNN). In the
  holographic model we construct a string worldsheet instanton which
  interpolates between a single and double string configuration, which
  determines the decay from one to two dual mesonic particles. The
  resulting probability for meson decay incorporates both the effects
  of finite meson size as well as back-reaction of the produced quarks
  on the QCD flux tube.  In the limit of very large strings the
  probability for a split reduces to the CNN formula. A byproduct of
  our analysis is the analysis of the moduli space of a generic double
  concentric Wilson loop with circles which are separated in the
  holographic direction of the confining background.  }

\begin{document}

\maketitle

\section{Introduction}

The holographic approach offers a framework to address some of the
most challenging questions in strongly coupled gauge theories in a
(semi) analytic way. While most of the work in the holographic
approach has taken place in the context of supersymmetric theories,
the expectation is that similar methods can be applied to study of
strongly coupled phenomena in QCD. At the moment the geometry which is
dual to QCD is not yet known. However, there are proposals for dual
geometries which capture various qualitative features of QCD.  One of
the most successful dual models is the Sakai-Sugimoto
model~\cite{Sakai:2005yt} which is special in the sense that it 
incorporates chiral symmetry breaking in the dual description.

In this paper we have used the Sakai-Sugimoto model to compute
probabilities for decays of mesonic particles, via breaking of flux
tubes. As this process is a strongly coupled phenomenon, its
computation in QCD is not easily performed.  Yet, knowing the
probability for a flux tube to break is crucial for understanding both
the decay widths of mesons and the hadronisation phase in high-energy
scattering processes.  A long time ago, a very successful
phenomenological model, the Lund fragmentation
model~\cite{Sjostrand:1982fn,Andersson:1983ia} was developed in order
to model hadronisation in event generators for high-energy
collisions. In this model, mesons are modelled by two (massive)
particles which are connected by a relativistic string, which models
the QCD flux tube. In a high-energy collision, the pair-produced quark
and antiquark move away from each other, with the colour string
stretching between them. As the string becomes longer, it eventually
snaps, producing a new quark-antiquark pair and so on, leading to a
shower of mesonic particles.

%For such a
%process a totall probability is given  by an area law
%\begin{eqnarray}
%  \label{lundareas}
%Probablity \sim e^{- b \, \, Area} 
%  \end{eqnarray}
%where Area is the area spanned by broken string worldsheets, and $b$
%is a probability per unit volume for a flux tube to break.  Probability $b$ is

The probability for a string to break at particular point was
``derived'' by Casher, Neuberger and Nussinov (CNN) in the early days
of QCD~\cite{Casher:1978wy}.  The formula was written down by making
an analogy with electromagnetism: the electric field in the Schwinger
formula was replaced with an (abelianised) chromoelectric field and
quarks were treated as free charged particles which are minimally
coupled to this field. While this model agrees qualitatively with
experimental data, it contains several free parameters which need to
be fixed by comparison with experiment.  A holographic approach may
potentially shed some light on the origin of these free
parameters.\footnote{We should note that the probe-brane
  approximation, in which back-reaction of the flavour brane is not
  taken into account, corresponds to the quenched approximation in
  QCD, in which dynamical features of quarks are neglected. One might
  thus say that in this approximation one cannot see any dynamical
  features of quarks, in particular one should not be able to see
  quark pair production as these correspond to~$N_f/N_c$
  corrections. However, generically the situation is more subtle than
  this. See in particular~\cite{Armoni:2008jy} which puts forward a
  proposal to compute signals of string breaking through the
  computation of the correlators of connected Wilson loops. The
  computation of~\cite{Armoni:2008jy} is in spirit very similar to
  what we do in the present paper.}

%  , as various explicit computations show. In some situations, it
%  is possible to set up a computation in the probe approximation which
%  is capable of capturing some features of dynamical quarks:
%  in~\cite{Aharony:2007uu} various properties of QCD at finite
%  chemical potential have been computed, and
%  in~\cite{Chakrabortty:2014kma,Semenoff:2011ng} Schwinger pair
%  production of quarks has been analysed.
  
The probability for a QCD string to break by producing a
quark-antiquark pair is also relevant when computing the lifetime of
mesons.  In the Sakai-Sugimoto model, large-spin mesons are modelled
by macroscopic, rotating, U-shaped strings with endpoints which are
stabilised from collapse by a centrifugal force and are constrained to
``move'' on probe D8-branes, see figure~\ref{backgroundprobe}. The
probability for such a string to split can be computed in two ways.
In our previous work~\cite{Peeters:2005fq,Sonnenschein:2017ylo} we
used a string bit model, in which we computed the probability for a
string to fluctuate in the holographic direction and hit the probe
brane. As it hits the probe, the string can split with some
probability.  The resulting decay width~$\Gamma$ was found to exhibit
exponential suppression in the masses of the pair-produced quarks and
linear dependence on the effective length of the QCD string flux
tube. Wave-function based approaches as in~\cite{Peeters:2005fq} are,
however, numerically hard to handle in the continuum limit, where the
number of string beads is large. In addition, the computations of
string fluctuations in~\cite{Peeters:2005fq} were for computational
reasons restricted to the near-wall region where the background metric
is linearised around flat space.

In order to improve on these points, we initiate in the present paper
an alternative, instanton approach to the study of holographic
breaking of the QCD flux tube. That is, we will construct a string
worldsheet instanton which interpolates between the unsplit and split
U-shaped mesonic strings.  As in~\cite{Peeters:2005fq}, we consider a
simplified system, which is represented by a hanging U-shaped string,
which does \emph{not} rotate but is prevented from collapse by a
Dirichlet boundary condition. Such a system is similar to the strings
used in the original Lund model, which was initially also applied to
non-rotating systems. The instanton configuration has the geometry of
a cylindrical surface, with circular boundaries which are concentric
in the field theory directions, and separated from each other in the
holographic direction of the dual geometry. A generic instanton
configuration would take into account the backreaction of the produced
quarks on the flux tube, through the bending of the flux tube in the
holographic direction. Our instanton describes the decay of a
finite-size flux tube and finite-volume mesonic particle in which the
endpoint quarks accelerate from each other.

The CNN formula does not take into account backreaction of the
pair-produced quarks and it also deals with an infinitely long flux
tube.  In the large volume limit, the QCD flux tube is much longer
than the radius at which the quarks are pair-produced and one expects
that the dynamics of the external quarks is decoupled from the string
breaking process. The probability is then fully determined by the
property of the tube and does not depend on the quarks in the original
meson. In order to compare our findings with the results of CNN, we
have therefore also investigated the large-volume limit of our result,
in which we indeed reproduce the simple exponential suppression of the
decay probability with the square of the quark mass, $\exp(-m_q^2/T)$,
where~$T$ is the tension of the string~\cite{Casher:1978wy}.
 
Our paper is organised as follows. In section~\ref{s:flat_breaking} we
first review the key features of the worldline derivation of the
Schwinger pair production formula~\cite{Affleck:1981bma} and its
generalisation to QCD~\cite{Casher:1978wy}. In
section~\ref{s:flat_instanton} we consider, as a warm-up exercise, the
old string model with massive endpoints in flat space, and construct
the instanton configuration which reproduces results
of~\cite{Casher:1978wy}. In section~\ref{s:ss_instanton} we construct
a similar instanton configuration in the Sakai-Sugimoto model and
compute from it the probability for meson decay. Our main findings and
open questions are discussed in the last section.

% A priory we are not guaranteed that will hold, since
%particle pair production of Schwinger is for IR free theory!, and we
%are looking here at the linearly growing potential.  Question is if pair
%production is probing the UV or IR of physics?

\section{QCD string breaking \`a la Schwinger in flat spacetime}
\label{s:flat_breaking}

It has been known for a long time that the presence of an external
electric field leads to the production of electrically charged
particle-antiparticle pairs~\cite{Schwinger:1951nm}. While the
original computation of Schwinger was done by perturbatively summing a
class of one-loop diagrams in quantum field theory, the same result
was later rederived in the worldline approach, by construction of a
worldline instanton~\cite{Affleck:1981bma}. The same worldline
instanton approach was also used to describe the production of
monopole/anti-monopole pairs in an external magnetic
field~\cite{Affleck:1981ag}.

In this section we will briefly review the basic derivation of the
Schwinger result for a production of particle-antiparticle pairs in
an external electric field, using the worldline instanton
approach~\cite{Affleck:1981bma}. We then review an application of this
formula to the pair production of quark-antiquark pairs inside the
QCD flux tube following the seminal work of Casher
et~al.~\cite{Casher:1978wy}.

\begin{figure}
\begin{center}
  \includegraphics[width=0.4\textwidth]{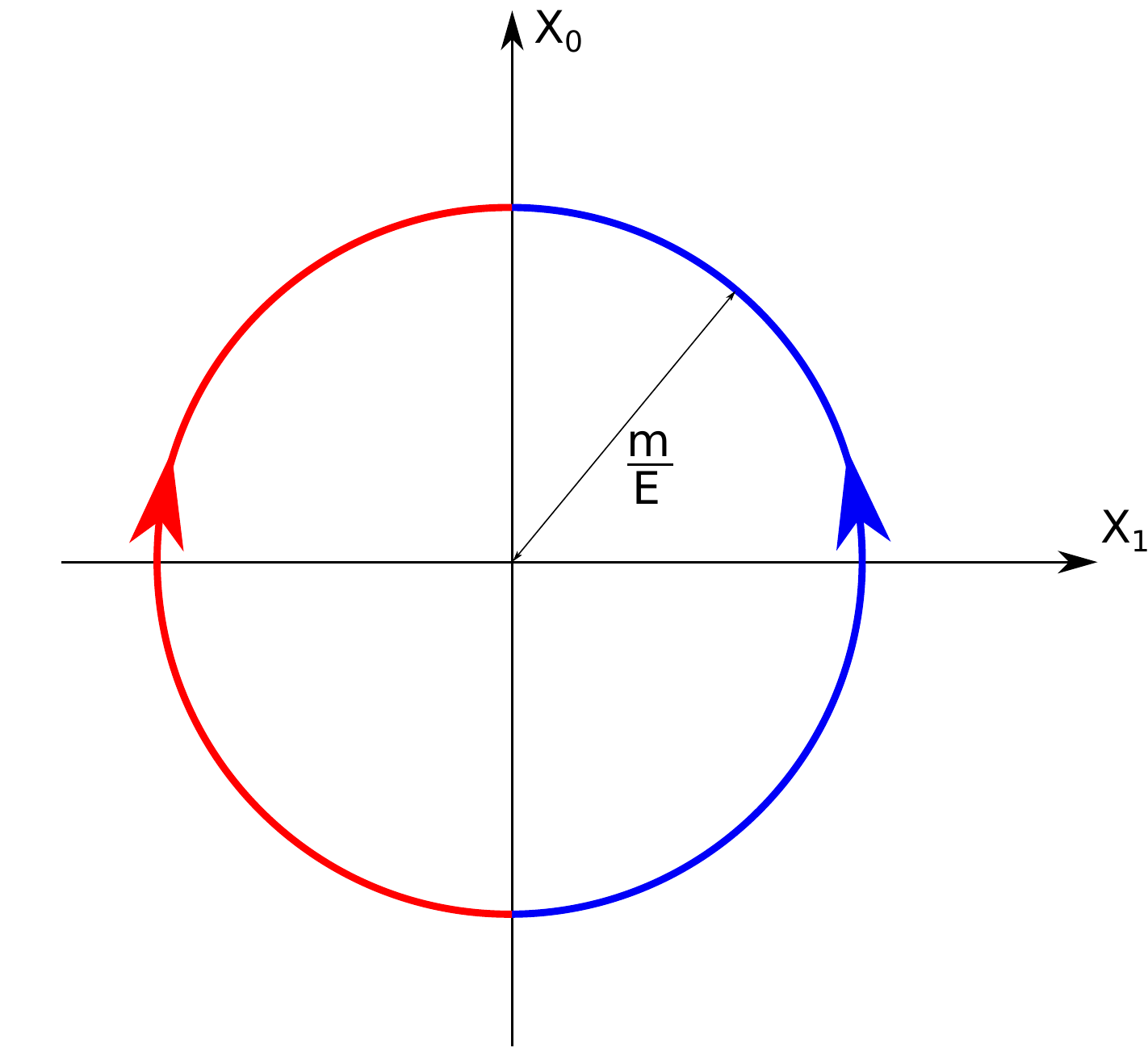}
\caption{\label{instantonoriginal} Worldline instanton for 
  particle-antiparticle pair production in an external electric
  field~$E$.}
\end{center}
\end{figure}

Assume that a non-vanishing electric field~$E$ is turned on in the
$X^1$ direction. In order to construct the worldline instanton
describing the production of a particle-antiparticle pair of
masses~$m$ and charge~$q$, one needs to consider the Wick rotated
system obtained by $\tau \rightarrow -i \tau$, $A_0 \rightarrow -i
A_0$ and solving the \emph{classical} equations of motion of a
particle in the Euclidean background with~$F_{01}=-iE$. The action of the particle is
given by
\begin{equation}
  \label{actionpp}
  S_{E} = \int\!{\rm d}\tau\, \bigg( m \sqrt{ \dot{X}^\nu \dot{X}_\nu}
  - i q A_{\nu} \dot{X}^\nu \bigg)\,.  
\end{equation}
It is not hard to see that the solution for the particle worldline is
given by
\begin{equation}
\label{instantonpp}
X^0(\tau) = R \cos(2 \pi n \tau) \, , \quad X^1(\tau) = R \sin(2 \pi n \tau) \, ,  \quad X^2 =0 \, , \quad X^3 = 0 \,,
\end{equation}
where $X^0$ is the Wick rotated target space time direction and~$R$ is
fixed in terms of~$E$ by the equation of motion, see below. We see
that the worldline instanton looks like a loop of radius~$R$. The
parameter~$n$ labels different instantons, and describes how many
times the particle ``winds'' around the loop. As usual, the particle
propagating ``backwards'' in the (Euclidean) time $X^0$ is interpreted
as an antiparticle. Hence, the left-hand side of the loop can be
interpreted as the worldline of the antiparticle, while the right-hand
side as the worldline of the particle, see
figure~\ref{instantonoriginal}.

Substituting the solution~\eqref{instantonpp} into the particle
action~\eqref{actionpp} and integrating over the worldline gives
\begin{equation}
  \label{e:worldlineaction}
  S_{\text{class}} =  2 \pi n R m -  \pi n q E R^2 \, .
\end{equation}
The extrema of the action will give classical solutions, and one finds
that the radius of the loop is fixed to be $R=m/(qE)$ for which the
action reduces to $S_{\text{class}}= \pi \frac{m^2}{qE} n$.
So the full loop~\eqref{instantonpp} describes a particle-antiparticle
pair which is produced at the Euclidean time $X^0=-R$, in which the
particles move away from each other.  Once the particle and
antiparticle go on-shell, i.e.~once they reach a distance $2m/(qE)$, one
can analytically continue the solution~\eqref{instantonpp} back to
Lorentzian time. The Lorentzian solution describes a pair of
particles accelerating away from each other with proper
acceleration \mbox{$a=m/(qE)$}.

Exponentiation of this Euclidean action $S_{\text{class}}$ with
winding~$n=1$ gives the most dominant contribution for the probability
of particle production in the saddle point approximation.  Looking at
the fluctuations around this classical path~\eqref{instantonpp}, and
summing over their contributions in the path
integral~\cite{Affleck:1981bma}, produces a pre-factor to the exponent
$e^{-S_{\text{class}}}$, and one obtains the celebrated Schwinger
formula for the probability of production of particles.  The
probability for the pair production of particles of spin half and
charge~$q$, per unit volume and per unit time, is given by
\cite{Schwinger:1951nm}
\begin{equation}
\label{exponent}
P_{\text{pp}} = \frac{E^2}{8 \pi^3} \sum_{n=1}^\infty \frac{1}{n^2} e^{-\frac{\pi m^2}{q |E|} n } \, .
\end{equation}

A long time ago, Casher et al.~\cite{Casher:1978wy} argued that the
Schwinger pair production formula~\eqref{exponent} can be directly
applied to QCD in order to derive a formula for the decay of
mesons. In their set up, Casher et al.~assumed that at the hadronic
energy scale of~$1$~GeV the quarks inside mesons can be treated as
Dirac particles with constituent masses $m$ and charge $q$. They also
assumed that at timescales which are short compared to the hadronic
timescale, mesons can be modelled as chromo-electric flux tubes
(``thick strings'') of universal thickness such that the
chromo-electric field can be treated as a classical, constant,
longitudinal \emph{abelian} field. Hence the process of meson decay
can be seen as Schwinger pair production of quark-antiquark pair, by
the (abelianised) QCD field.

The flux tube is parametrised by the radius $r_t$, the ``abelianised''
QCD field strength $\mathcal{E}_t$ and the gauge coupling $g$ which is
related to the charge of the quarks $q$ as $q=g/2$. It has been argued
that the reason for the factor $1/2$ between $g$ and $q$ is the fact
that quarks couple to the gauge field through the $SU(3)$ generators
$\lambda^a/2$.  The energy per unit length stored in the tube is the
\emph{effective} tube (string) tension and is given by
\begin{equation}
\gamma_{\text{QCD}} = \frac{1}{2 \pi \alpha'}= \frac{1}{2}\mathcal{E}_t^2 \pi r_t^2  \, ,
\end{equation}
$\gamma_{\text{QCD}} \sim 0.177 (\text{GeV})^2$ and the radius of the
tube is $r_t \sim 2.5 \text{GeV}^{-1}$.  On the other hand, using the
(abelian) Gauss law and the fact that flux lines are non-vanishing only
between the quarks (like for a capacitor), one has $\mathcal{E}_t
\pi r_t^2 = q = g/2$ which implies that the effective tension of the
flux tube is
\begin{equation}
\label{effectivestring}
\gamma_{\text{QCD}} =  \frac{1}{2}\mathcal{E}_t q \, .
\end{equation}
Hence the Schwinger formula~\eqref{exponent} can be rewritten in terms
of the natural QCD variables as
\begin{equation}
\label{QCDSchwinger}
P_{\text{QCD}} = \frac{\gamma_{\text{QCD}}}{ \pi^3} \sum_{n=1}^\infty \frac{1}{n^2} e^{-\frac{\pi m_{q}^2 }{ 2 \gamma_{\text{QCD}}} n } \, .
\end{equation}
We should comment at this stage that the factor of~$1/2$ in the
exponent is a consequence of the fact that the QCD field in the flux
tube has been treated as an abelian field. In the more recent
paper~\cite{Nayak:2005pf}, a proper generalisation of the Schwinger
formula to a non-abelian field has been derived and the production
rate has been shown to depend on two independent Casimir gauge
invariants~$E^aE^a$ and~$d_{abc}E^a E^b E^c$.  In what follows we will
see that the holographic model reproduces the exponential dependence
of the production rate in~\eqref{QCDSchwinger}, up to this numerical
factor.

From the formula (\ref{QCDSchwinger}), the probability for a meson to
decay after time $t$, measured in the meson rest frame, is $1-
e^{-V_4(t) P_{QCD}}$ where $V_4(t)$ is the four-volume spanned by the
system until time~$t$. For a meson which is modelled by a rotating
flux tube of lenght~$L$, this volume is $V_4(t) = \pi r_t^2 L
t$. Therefore, the decay width (probability per unit time) is $\Gamma=
\pi r_t^2 L P_{\text{QCD}}$. Because the meson mass is $M = \pi
\gamma_{\text{QCD}} L$, one finds that the ratio of decay width
$\Gamma$ and the meson mass $M$ is independent of the effective length
of the string,
\begin{equation}
  \left(\frac{\Gamma}{M} \right)_{\text{rot}} = \frac{2 r_t^2}{\gamma_{\text{QCD}}}
  P_{\text{QCD}}\,.
\end{equation}
Similarly, modelling mesons as one-dimensional oscillators implies
$(\Gamma/M)_{\text{osc}} = \pi/4(\Gamma/M)_{\text{rot}}$, so that the
ratio is independent of the effective size of the sysem. At the
moment, the experimental data on meson decays do not agree with this
prediction for lighter mesons (see the discussion
in~\cite{Peeters:2005fq,Sonnenschein:2017ylo}), while for high-spin
mesons, where one would expect this model to work bettter, the data
are not accurate enough to confirm or reject such a prediction.

\section{Worldsheet instanton in flat space and string splitting}
\label{s:flat_instanton}

While our main goal is to study meson decays in the holographic
setup, we will as a warm up exercise first consider the process of
meson decay in flat space without using the analogy with the Schwinger
formula. This will provide an alternative, new derivation of the
formula~\eqref{QCDSchwinger} which, to the best of our knowledge, has
so far not been presented elsewhere.

In order to model mesons, including their flux tube as well as
endpoint quarks, we will use an action for the relativistic string
suplemented with two massive particles which are attached to the
string endpoints. This is the ``old'' string model, as discussed and
reviewed in~\cite{Barbashov:1990ce}.  We want to find the Euclidean
worldsheet configuration which interpolates between the unsplit and
split string with massive end points. After performing a Wick rotation
in the target and worldsheet space-time, the string action becomes
\footnote{Note that we are using a mostly-plus Lorentzian metric.}
\begin{equation}
  \label{actionflat}
  \begin{aligned}
    S &=  \gamma \int_{\tau_1}^{\tau_2}\!{\rm d} \tau\, \int_{0}^{\pi} d\sigma \sqrt{-(\dot{X} \cdot X')^2 + \dot{X}^2 X'^2 }\\[1ex]
&\qquad\qquad    + m  \int_{ \tau_1}^{ \tau_2}\! {\rm d}\tau \bigg(\sqrt{\dot{X}^{2}(\tau , \sigma=0)} + \sqrt{\dot{X}^2(\tau, \sigma=\pi)}\bigg) \\[1ex]
    &\equiv S_{\text{bulk}} + S_{\partial, \sigma=0} + S_{\partial, \sigma=\pi }\,.
  \end{aligned}
\end{equation}
Here $X \cdot X \equiv X^{\mu} X^{\nu} g_{\mu\nu}(X)$ and
$g_{\mu\nu}(X)$ denotes the Euclidean metric in the target space,
which for us at this stage is just a flat metric, $g_{\mu\nu} =
\delta_{\mu\nu}$. The tension of the string is denoted with $\gamma=
1/(2\pi \alpha')$ and $m$ is the mass of the particles attached to the
string endpoints.

We are interested in finding a Euclidean, two-dimensional string
configuration which interpolates between a single and a double
string. The initial string had only two quarks at its endpoints.  In
the fully dynamical string model, the position of these outer quarks
is fixed by the total angular momentum of the meson, which prevents
strings from collapsing, see for example~\cite{Barbashov:1990ce}. To
simplify the discussion, we will in this paper take quarks in the
original meson to satisfy Diriclet boundary conditions and confine
them to move on a line in the Euclidean target space.

\begin{figure}
\begin{center}  
\includegraphics[width=0.8\textwidth]{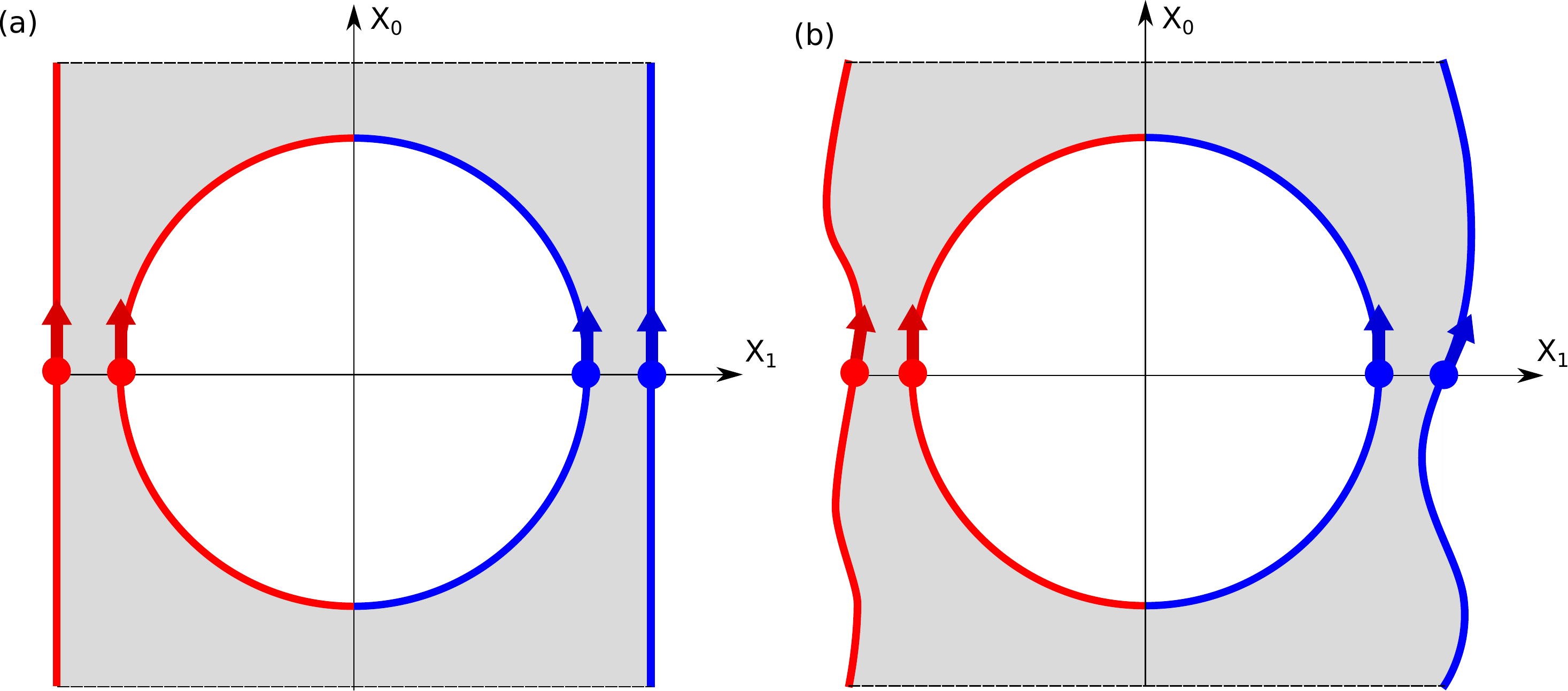}
\caption{\label{generalisedinstanton}Flat space instantons, describing
  the breaking of the string flux tube. The instanton on the left
  describes breaking of the flux tube where the external quarks are at
  a fixed distance, while the instanton on the right depicts external quarks
  which ``move'' on arbitrary paths.}
\end{center}
\end{figure}

At some point in the Euclidean ``time'', a pair of massive quarks is
pair-produced in the interior of the string. These particles represent
new ``internal'' endpoints of the string, which can move freely,
i.e.~satisfy Neumann boundary conditions, see
figure~\ref{generalisedinstanton}. In order to account for the
pair-produced quarks, one needs to modify the
action~\eqref{actionflat} by adding to it the worldline action of the
pair-produced quarks. Adding this extra term in the action is in
spirit the same as what one does in order to describe the
pair-production of the charged particle-antiparticle pair in the
external electric field using the instanton approach in the world-line
formalism, see e.g.~\cite{Semenoff:2011ng}. The main difference is
that the role of the electric field is now played by the tension of
the split string, which pulls apart the pair-produced particles.  For
simplicity, we will also assume that the particles have no transverse
momentum, so that the whole process of string splitting is planar,
i.e.~that both in- and outgoing strings are in the same
two-dimensional plane.

As the variation of the action~\eqref{actionflat} leads to bulk and
boundary equations of motion we have to make sure that both are
satisfied. In the two-dimensional target space, the bulk equations of
motion are always satisfied, thanks to reparametrisation invariance of
the action. So we just need to make sure that the boundary equations
of motion for the Neumann boundary conditions hold.  Note that the
boundary equations of motion receive nontrivial contribution from the
surface terms of the bulk part of the action,
 \begin{equation}
 \label{equationNeumann}  
\frac{\partial\,\,\,}{\partial\tau}\left(\frac{\partial
  \mathcal{L}_{\partial, \sigma_B}}{\partial
  \dot{X}^0}\right)-\frac{\partial \mathcal{L}_{\text{bulk}}}{\partial
  {X'}^0}\Bigg|_{\sigma_B} = 0\,,
\end{equation}
where $\sigma_B$ is the position of the boundary (or boundaries) of
the string for which Neumann conditions are imposed.

Before the quarks were pair-produced, the string was straight
and stretching between $(0,2 L)$ in the target space. To describe this
string configuration we choose the parametrisation
\begin{equation}
\label{background}
X^0 = \tau\,, \quad \quad X^1 = 2 L \frac{\sigma}{\pi} \,,  \quad\quad \sigma \in [0, \pi\rangle\,.
\end{equation}
The instanton configuration for a splittting string is plotted in
figure~\ref{generalisedinstanton}, and is given by
\begin{equation}
  \label{solusplit}
  \begin{aligned}
   X_L^0(\tau,\sigma)&=X_R^0(\tau,\sigma)=\tau\,, \\[1ex]
	X_L^1(\tau,\sigma)&=x_L(\tau,\sigma)=-\frac{\sigma}{\pi}\left(\sqrt{-\tau^2+
     \kappa^2}- a \right)\,, & \sigma \in [0, \pi\rangle \,, \\[1ex]
	X_R^1(\tau,\sigma)&=x_R(\tau,\sigma)=\left(1-\frac{\sigma}{\pi}\right)\left(\sqrt{-\tau^2+\kappa^2}+
   a \right)+2 a \frac{\sigma}{\pi}\,, &\sigma \in [0, \pi\rangle\,,
  \end{aligned}
\end{equation}
where $\kappa$ and $a$ are arbitrary constants, and $X_L$ and $X_R$
describe left and right half of the instanton (the red and blue areas
in figure~\ref{generalisedinstanton}). Note that while we have written
the solution piece-wise, the two ``sides'' of the instanton, $X_L$ and
$X_R$ are glued in a smooth way. The solution above is a Euclidean
version of a solution found in~\cite{Bardeen:1975gx}.

It is easy to see that the ansatz~\eqref{solusplit} satisfies the
Neumann boundary equations of motions~\eqref{equationNeumann},
\begin{equation}
  \label{fullinstanton}
	-m\frac{\partial\,\,\,}{\partial\tau}\left(\frac{\dot{X}_L^\mu(\tau, \sigma=\pi)}{\sqrt{1+\dot{x}_L^2(\tau,\sigma=\pi)}}\right)+\gamma\left(-\dot{x}_L\dot{X}_L^\mu+\left(\frac{1+\dot{x}_L^2}{x_L'}\right)X_L'^\mu\right)\bigg |_{\sigma=\pi}=0\,,
\end{equation}
(where $\mu=0,1$) provided that $\kappa= m/\gamma$. A similar
expression holds for the right-hand piece~$X_R^\mu$. Note also that
the solution~\eqref{solusplit}, where the outer quarks are moving on
straight lines, can be generalised to a solution for which the
endpoints do not follow straight lines but move on arbitrary curves
$f_{L}(\tau)$ and $f_R(\tau)$, see figure~\ref{generalisedinstanton}b.
\begin{equation}
  \label{solusplitV2}
  \begin{aligned}
      X_L^0(\tau,\sigma)&=X_R^0(\tau,\sigma)=\tau\,, \\[1ex]
	   X_L^1(\tau,\sigma)&=x_L(\tau,\sigma)=-\frac{\sigma}{\pi}\left(\sqrt{-\tau^2+
        \kappa^2}- a \right) + f_L(\tau) \left( 1- \frac{\sigma}{\pi}
      \right)\,, \quad&  \sigma & \in (0, \pi)\,,  \\[1ex]
      X_R^1(\tau,\sigma)&=x_R(\tau,\sigma)=\left(1-\frac{\sigma}{\pi}\right)\left(\sqrt{-\tau^2+\kappa^2}+ a \right) + f_R(\tau) \frac{\sigma}{\pi} \,,  \quad& \sigma &\in (0, \pi) \,,\\
  \end{aligned}
\end{equation}
with $\kappa = m/\gamma$ as above. We therefore see that in whichever
way the outer quarks move, the dynamics of the inner free (Neumann) quarks
is unaffected. The ``motion'' of the inner quarks is always circular,
with a radius of curvature which is determined by the ratio of
the particle mass and the string tension which pulls the produced quarks.

In order to evaluate the probability for a single event of particle
pair production, we need to evaluate the action of the instanton. The
string configuration (\ref{solusplit}) describes the process of pair
production inside the string, as well as the propagation of the
outer, background quarks. Hence in order to isolate the part which
describes particle production, we need to subtract the contribution
corresponding to the background. In other words, the quantity of
interest which gives us the probability for the particle production is
\begin{multline}
\label{actionproduction}  
  S_{\text{pp}} = S_{\text{full}} - S_{\text{background}} =
  m(\mathrm{Circumference}\,\,\mathrm{of}\,\,\mathrm{circle})-\gamma(\mathrm{Area}\,\,\mathrm{of}\,\,\mathrm{circle})
  \\[1ex] = m(2\pi \kappa)-\gamma(\pi \kappa^2)= \frac{\pi
    m^2}{\gamma}\,.
\end{multline}
Here $S_{\text{background}}$ is the action of the background
configuration with no particle production~\eqref{background},
$S_{\text{full}}$ is the action of the instanton
configuration~\eqref{fullinstanton} and~$\gamma$ is the string
tension. It is easy to see that the same result is obtained with the
more general solution~\eqref{solusplitV2}, where the outer quarks move
on arbitrary, non-straight paths. In other words, the dynamics of the
external quarks is fully decoupled from the production process inside
the flux tube.

We thus find that the probability for a single pair production event
to happen is given by
\begin{equation}
\label{particlestuff}
  P_{\text{PP}} = e^{- S_{\text{pp}}} = e^{-\frac{\pi m_q^2}{\gamma}} \, ,
\end{equation}
We see that this is the as the contribution of a single instanton in
the CNN formula (\ref{QCDSchwinger}), up to a numerical factor of a
half.  So we see that in the process of string splitting, the string
tension has the same role as the electric field in the Schwinger
process, that is, it pulls the produced particles away from each
other. However, note again that there is a difference with respect to
the Schwinger process, as quarks couple to the string endpoints in a
different way than described by minimal coupling to electromagnetism
(as used in the CNN approach).

The position of the instanton can be at any arbitrary point on the
string worlsheet, as long as the instanton is not too close to the
boundary of the string worldsheet, so that size of the instanton
circle ``fits'' into the string worldsheet. One can therefore compute
the probability per unit time $\Gamma$ from the the probability per
unit volume and unit time~\eqref{particlestuff} as $\Gamma
=P_{\text{pp}}L$ where $L$ is the string length. On the other hand,
the mass of the initial mesonic ``particle'' is $M = \gamma L + 2m$,
where $m$ is the mass of the initial quark pairs. In the limit of long
strings ($L\gg m/\gamma$) one can approximate $M\approx \gamma L$ and
also ignore subtleties related to whether the instanton fits into the
worldsheet or not. In this limit one recovers the results from the
previous section as well as from the holographic computation
of~\cite{Peeters:2005fq}, namely that $\Gamma/M$ is a constant for all
mesonic particles.

In summary, we have constructed a flat space instanton configuration
which describes splitting of the open relativistic string with the
massive endpoints, into two strings with massive endpoints. The
probability for such a process to occur is the same as the probability
for pair production of charged massive particles in an external
electric field of a strength which is proportional to the tension of
the string.

\bigskip

\section{String splitting in the Sakai-Sugimoto holographic model}
\label{s:ss_instanton}

Our discussion of splitting strings was so far done in flat space
using the old string model of mesons. In holographic models of QCD,
like the Sakai-Sugimoto model, mesons are incorporated by adding one
or more flavour D8 probe branes in the holographic background which is
dual to the confining, pure glue theory at strong
coupling~\cite{Sakai:2005yt}.  In this setup the mesons appear either
as light fluctuations of the probe D8-brane in the supergravity (DBI)
approximation or as semi-classical string configurations of
relativistic strings whose worldsheets end on the probe brane.  Light
DBI excitations of the probe brane describe mesons up to spin one,
while higher-spin mesons are described by semi-classical strings. In
what follows we will focus on the phenomenologically more interesting
case of higher-spin mesons and their decay by constructing string
worldsheet instanton configurations in the holographic background.

\subsection{Review of high-spin mesons in the Sakai-Sugimoto model}

The background geometry in which probe D8-branes and large strings are
embedded is given by
\begin{multline}
  \label{backgroundgeometry}
   {\rm d}s^2 = \left(\frac{u}{R_{D_4}}\right)^{3/2}\left(-{\rm d} t^2+ \delta_{ij}{\rm d} x^i {\rm d} x^j + f_\Lambda(u){\rm d} x_4^2 \right) +  \left(\frac{R_{D_4}}{u}\right)^{3/2} \bigg(\frac{{\rm d} u^2 }{f_{\Lambda}(u)}   + u^2  {\rm d} \Omega_4 \bigg)\,, \\[1ex]
f_\Lambda(u) = 1-\frac{{u_\Lambda}^3}{u^3}\,,  \quad \quad  i=1,2,3  \, , 
\end{multline}
where ${\rm d}\Omega_4$ is the metric on a round four sphere. There is
also a non-constant dilaton and a RR four-form field
strength,
\begin{eqnarray}
e^\phi = g_s \left( \frac{u}{R_{D_{4}}} \right)^{3/4} \quad \quad F_4 = \frac{2 \pi N_c}{V_4} \epsilon_4 \, .
\end{eqnarray}
Here $R_{D_4}^3= \pi g_s N_c l_s^3$, $g_s$ is the string coupling and
$l_s^2=\alpha'$ is the string length. Here $u$ is the ``holographic''
direction, which is bounded from below by $u \geq u_{\Lambda}$. The
world volume, non-holographic, directions in which the gauge theory
lives are $t,x_1,x_2,x_3$. One of the main properties of this
background is the cigar-like submanifold spanned by the periodic
coordinate~$x_4$ and the holographic direction $u$.  The tip of the
cigar is positioned at~$u=u_{\Lambda}$ where the $x_4$~circle
(smoothly) shrinks to zero size.\footnote{In order to ensure that tip
  of the cigar is non-singular, the periodicity of~$x_4$ has to be
  \begin{eqnarray}
    \delta x_4 = \frac{4
      \pi}{3}\left(\frac{R_{D_4}^3}{u_{\Lambda}}\right)^{1/2} \equiv 2
    \pi R \, .
    \end{eqnarray}
} One usually refers to the region near $u_{\Lambda}$, as the
``wall''.

\begin{figure}
  \begin{center}  
    \includegraphics[width=0.7\textwidth]{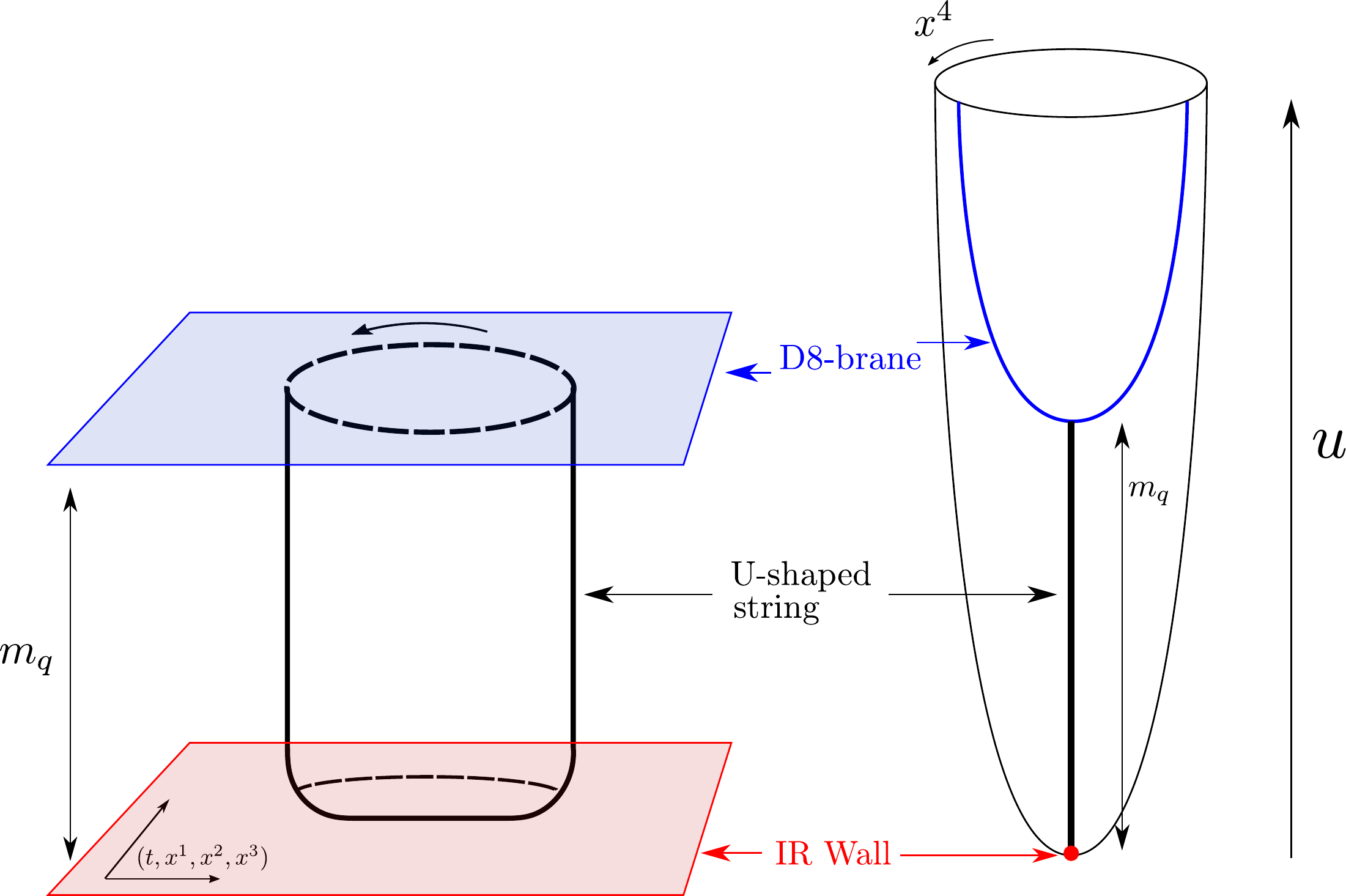}
  \end{center}
  \caption{\label{backgroundprobe}Sakai-Sugimoto background with the
    probe D8-brane and the U-shaped, mesonic string which hangs in the
    holographic direction towards the wall at~$u=u_{\Lambda}$.}
\end{figure}

In order to incorporate quark/flavour degrees of freedom to this pure
glue theory, one needs to place probe flavour $D8$ brane(s) in this
geometry.  There are different ways in which one can embed flavour
$D8$ branes in this background. For us, the relevant embedding of the
probe flavour $D8$ brane is the one in which $D8$ brane fills out all
directions except the cigar $(u,x_4)$ submanifold.  In the cigar
submanifold, the flavour $D8$ brane has a U-shape, see
figure~\ref{backgroundprobe}, with the tip of the probe brane which is
at some distance $m_q$ from the wall $u_{\Lambda}$,
see~\cite{Sakai:2005yt}.  In principle this parameter~$m_q$ is a free
parameter for the embedding of the brane, and can be changed by
changing the asymptotic separation between the endpoints of the probe,
see figure \ref{backgroundprobe}.

Large spin mesons correspond to rotating strings, whose endpoints are
fixed on the flavour D8-brane. The strings are prevented from
collapsing by a centrifugal force~\cite{Kruczenski:2004me}. As the
spin of the string is increased, the distance between the string
endpoints increases as well, i.e.~the string becomes larger and its
worldsheet becomes and more and more U-shaped. The two ``vertical''
parts of the string stretch almost vertically from the probe brane to
the wall and the horizontal part of the string stretches almost
parallel to the wall, see figure~\ref{backgroundprobe}. It was shown
in~\cite{Kruczenski:2004me} that this string configuration is
holographically equivalent to the system of two quarks which are
connected by a flux tube, i.e.~to the same model we have analysed in
the previous section. By analysing the mass of this string
configuration, it was shown that the vertical parts of the string
correspond the (bare) quark masses of the meson, while the horizontal
part of the string corresponds to the energy stored in the QCD flux
tube.

In order to model a system with different quark masses for the two
quarks, one needs to introduce more than one flavour D8-brane, each
hanging at different distance from the wall. The positions of these
probes in the holographic diretion specify different quark masses. A
meson with different quark masses is then a string with endpoints
ending on these two different flavour D8-branes.

We would now like to study the decay of such a string configuration.
The hanging string is subject to quantum fluctuations and when a part
of the string worldsheet touches one of the flavour branes it can
split and attach new endpoints to that flavour brane. The probability
for a string to touch the flavour brane due to quantum fluctuations
was computed in~\cite{Peeters:2005fq} by constructing the string
wave-function using a string bead model for a discretised string
worldsheet.

Our approach here will be different. We will here construct a
configuration of the Euclidean worldsheet, which interpolates between
the single and double U-shaped strings, that is, a worldsheet
instanton.

\subsection{String worldsheet instanton}

Our main goal is to holographically compute the probability per unit
volume and time for a QCD flux tube to break. As in our previous
analysis~\cite{Peeters:2005fq} and as in the flat space construction
from the previous section, we simplify the problem by looking at a
U-shaped string with endpoints which are ``forced by hand'' to follow
a circular path of some radius. Imposing these boundary conditions is
not unreasonable, as in the Lorentzian picture these correspond to
quarks which accelerate away from each other with constant
acceleration, as it happens in the hadronisation phase in high-energy
scattering processes.\footnote{\label{f:nonsymmetric} A better
  description of meson decays would involve string surfaces with
  straight external boundaries and a circular inner boundary. However,
  this breaks the rotational symmetry and the computation is thus
  numerically much more involved. In what follows we will nevertheless
  often loosely refer to the size of the outer radius as ``the size of
  the meson''. We intend to return to the non-symmetrical problem in a
  followup paper.}

Generically the string breaking process will be sensitive to the
precise boundary conditions one imposes for the external
quarks. However, one would expect that there is a limit in which the
exact dynamics of the external quarks decouples from the breaking
process (a sort of large volume limit), so that the quark production
process in this limit can be treated as a Schwinger process in a
constant external field, like in~\cite{Casher:1978wy}. As it is a
priori not clear what this limit is or whether it exists in our setup,
our approach will be to first construct the general solution and
compute the decay probability for an arbitrary mesonic particle, and
then see if there is a limit in which this probability reduces to the
Schwinger probability.

A string can break only at the point where the interior of the world
sheet touches the probe D8-brane. In real time such a situation
happens because under quantum fluctuations, parts of the string
worldsheet touch the probe brane.\footnote{Here we are considering
  breaking of an open string into two other open strings. Note
  however, that there is an alternative channel of decay in which the
  open string radiates closed loops. In this process one does not
  require that the string worldsheet touches the probe, but rather
  any self intersection of the string worldsheet can lead to the
  emission of closed strings. However, this process does not describe
  the process of a meson decay into two mesons, but a decay of a meson
  into another meson plus a glueball. Such a process is suppressed by
  additional powers of~$g_s$ which suppresses open to closed string
  amplitudes with respect to open to open string amplitudes, and it
  will not be analysed here. } In the Euclidean setup, in order to
construct the instanton configuration for a splitting string, one
needs to start with the string worldsheet which is ``pinned'' to the
D8 probe at some internal worldsheet point. So we impose Dirichlet
boundary conditions in~$u$ and~$x_4$ directions both for the string
endpoints and for the ``pinning point''. Once the string has split at
the pinning point, the newly generated string endpoints are free to
``move'' in the D8~worldvolume directions ($x_0,x_1,x_2,x_3$ and $S^4$)
freely, i.e.~they satisfy Neumann boundary conditions.

In order to construct worldsheet instanton, we need to solve the
string equations of motion in the Wick rotated
background~\eqref{backgroundgeometry}. It will be convenient to change
to background coordinates as follows,
\begin{equation}
  z = \frac{1}{u}\,, \quad \quad z_{D_4} = \frac{1}{R_{D_4}}\,,  \quad \quad  z_{\Lambda} = \frac{1}{u_{\Lambda}} \, ,
\end{equation}
which turns the metric into
\begin{equation}
  \label{backgroundgeometryV2}
  \begin{aligned}
   {\rm d}s^2 = \left(\frac{z_{D_4}}{z}\right)^{3/2}\bigg({\rm d}
   x_0^2 + {\rm d}\rho^2 + \rho^2 {\rm d}\theta^2 &+ \rho^2 \sin \theta
   {\rm d} \phi^2  + f_\Lambda(z){\rm d} x_4^2 \bigg) \\[1ex]
   &+  \frac{1}{z_{D_4}^{3/2} z^{5/2}} \bigg(f_{\Lambda}^{-1}(z) {\rm d} z^2  + z^2  {\rm d} \Omega_4 \bigg)\,,\\[1ex]
f_\Lambda(z) &\equiv 1-\frac{z^3}{z_{\Lambda}^3}  \quad \quad  i=1,2,3  \, , 
  \end{aligned}
\end{equation}
and $0 \leq z \leq z_{\Lambda}$. We have Wick-rotated time and we have
also introduced spherical coordinates in the $(x_1,x_2,x_3)$ directions.

The string worldsheet extends in the radial direction~$z$, has
cylindrical symmetry in the worldvolume directions, and hangs from a
fixed position in the~$x_4$ direction, which is at the tip of the
D8-probe. A standard coordinate choice on the worldsheet is the
static gauge, in which one makes the following ansatz for the string
worldsheet,
\begin{equation}
\label{parametrisation}
  z = \sigma\,, \quad \quad  \rho= \rho(z)\,, \quad \quad 
  \theta = \frac{\pi}{2}\,, \quad \quad \phi = \tau \, . 
\end{equation} 
Plugging this into the string action one gets
\begin{equation}
  {\mathcal L} = \gamma \sqrt{- ({\dot X} \cdot X')^2 + (X')^2 {\dot X}^2}
  = \gamma \frac{\rho}{z^{3/2}} \sqrt{z_{D4}^3 \rho'^2 + \frac{1}{z f_{\Lambda}(z)} }\,,
\end{equation}
which leads to the equations of motion
\begin{multline}
\label{equationsofmotionV1}
    2z_{\Lambda}^3 \bigg(z_{\Lambda}^3 + z_{D_4}^3z(-z^3 + z_{\Lambda}^3)\rho'^2 \bigg) + \\[1ex]
 + z_{D_4}^3 \rho \bigg( z_{\Lambda}^3(z^3 + 2 z_{\Lambda}^3)\rho' + 3 z_{D_4}^3 z (z^3 - z_{\Lambda}^3)^2 \rho'^3 + 2 z z_{\Lambda}^3 (z^3 -z_{\Lambda}^3)\rho'' \bigg) = 0 \, .
\end{multline}

The above choice for the worldsheet coordinate is, however, not very
good when constructing numerical solutions. The U-shaped strings we
are after have, in this system, parts in which either the~$z'$ or
$\rho'$ derivatives are large. In fact, because of the combination of
almost vertical and almost horizontal segments, no single coordinate
system turned out to be particularly well-suited to finding reliable
solutions in all regions of the parameter space which we have
explored. We have therefore used a numerical solution method which
automatically switches between three different coordinate systems on
the worldsheet ($\sigma=z$, $\sigma = z+\rho$ and $\sigma=\rho$) so
as to keep the solution regular.

% of the fact that
% 
% 
% Near the wall, at the bottom of the string worldsheet, the string is
% (almost) parallel to the brane. Hence the parametrisation
% (\ref{parametrisation}) is not convenient for this part of the string
% worldsheet. In order to do numerical integration of the equations of
% motion in that region, we have used alternative parametrisation
% \begin{eqnarray}
% \label{parametrisationalternative}
%   \rho &=& \sigma \quad \quad z = z(\rho) \quad \quad \theta =
%   \frac{\pi}{2} \quad \quad \phi = \tau \, \, .
% \end{eqnarray} 
% With this parametrisation, the string  action becomes
% \begin{eqnarray}
% {\mathcal L} = \gamma \frac{\rho}{z^2} \sqrt{z_{D4}^2  +  \frac{z'^2}{ f_{\Lambda}(z)} }
% \end{eqnarray}
% with the equations of motion
% \begin{eqnarray}
% \label{equationsofmotionV2}  
%  && - 2 \rho z_{\Lambda}^8 \bigg(z'^2 + 1 -  z_{D_4}^8 z(-z^3 + z_{\Lambda}^3)\rho'^2 \bigg) + \nonumber \\
% && \quad \quad \quad \quad \, \,    + z_{D_4}^3 \rho \bigg( z_{\Lambda}^3(z^3 + 2 z_{\Lambda}^3)\rho' + 3 z_{D_4}^3 z (z^3 - z_{\Lambda}^3)^2 \rho'^3 + 2 z z_{\Lambda}^3 (z^3 -z_{\Lambda}^3)\rho'' \bigg) = 0 \, .
% \end{eqnarray}

\begin{figure}
  \begin{center}  
    \includegraphics[width=0.4\textwidth]{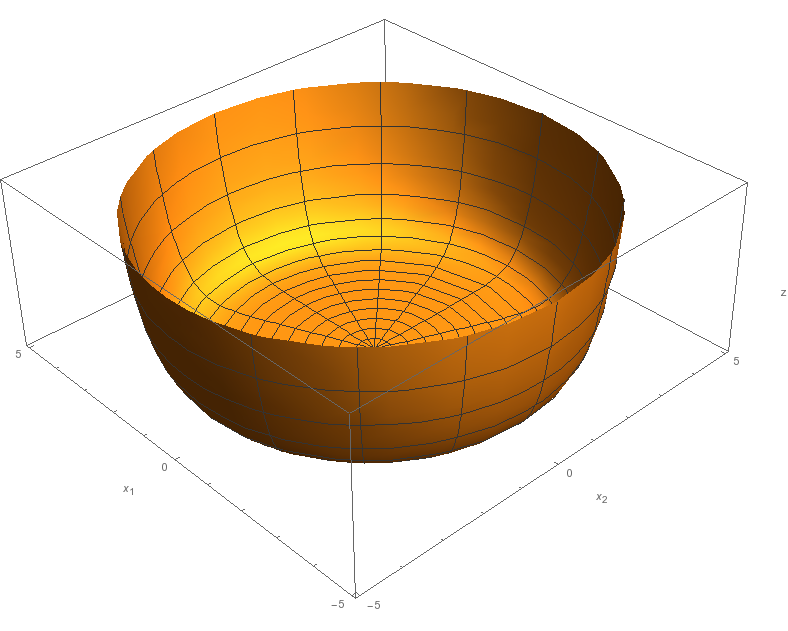}\quad
    \includegraphics[width=0.5\textwidth]{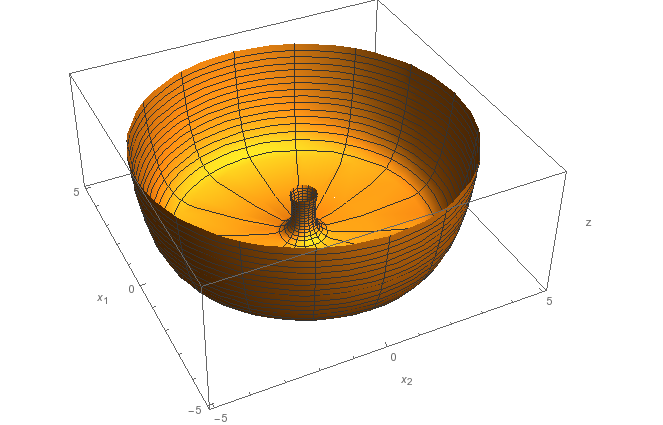}
  \end{center}
  \caption{\label{VariousSolutions} Typical single loop solution {(\bf
      I)}, left and a double loop solution {(\bf II)}, right.}
\end{figure}

The equations of motion~\eqref{equationsofmotionV1} admit two types of
solutions which have different topologies and satisfy different
boundary conditions at the string endpoints. The first solution
corresponds to the the single U-shaped string and it describes the
(original) quark and antiquark which are forced to ``move'' on a
circular orbit, and are connected by a flux tube. If one was to Wick
rotate this configuration to Lorentzian time, it would correspond to a
quark and antiquark which accelerate away from each other, while being
connected by a flux tube.  We will refer to this solution as
solution~({\bf I)}; see the left-hand side plot in
figure~\ref{VariousSolutions}.

The second solution is a string with two disconnected boundaries,
which the describes process of flux tube breaking. The outer boundary
of the string is forced by a Dirichlet boundary condition to be on a
circle of a fixed radius~$R_{2}$. The inner boundary is forced to be
on a particular D8 probe (with Dirchlet boundary conditions in
the~$x_4$ and~$z$ directions), but the internal ends of this string
are free to ``move'' arbitrarily along the D8 probe (Neumann boundary
conditions). The physical reason why we impose ``free'' Neumann
boundary conditions on the inner edge of the string is that this part
of the worldsheet corresponds to the pair-produced quarks which ``move''
only under the influence of the flux tube, and are not coupled to any
external source.  We will refer to this solution as solution~({\bf
  II}), see right-hand side plot in figure~\ref{VariousSolutions}.

It is not hard to see that if one imposes a Dirchlet boundary condition
in the~$z$ direction, then the inner boundary of the hanging string
has to ends orthogonally on the D8-brane worldvolume. Namely, in the
gauge~$z=\sigma$, the Neumann boundary condition in the~$\rho$ direction
yields
\begin{equation}
\label{orthogonal}  
 \frac{\partial {\mathcal L}}{\partial \rho'} = \gamma
 \left(\frac{z_{D4}}{z}\right)^\frac{3}{2} \bigg(z_{D_4}^3 \rho'^2 +
 (z f_{\Lambda})^{-1} \bigg)^{- \frac{1}{2}} \rho \rho' = 0 \quad \Rightarrow \quad  \frac{{\rm d} \rho}{{\rm d} z} = 0 \, ,  
\end{equation}
i.e.~the string hangs orthogonally from the D8 probe.

For both string configurations, the constituent quark masses are given
by~\cite{Kinar:1998vq}
\begin{equation}
\label{quarkmasses}
m_{Q}  = \frac{1}{2 \pi \alpha'} \int_{z_\Lambda}^{z_{m_Q}} {\rm d}  z \, \sqrt{g_{zz} g_{00} } =  \frac{1}{2 \pi \alpha'}  \int_{z_\Lambda}^{z_{m_Q}} {\rm d}  z \, \frac{1}{f_{\Lambda}^{1/2} z^{2}} \, ,
\end{equation}
which is just the proper distance of a string hanging from the tip of
the probe D8-brane, $z_{m_Q}$ to the IR wall at~$z_{\Lambda}$.  Note
that if there is more than one probe D8-brane, which each ends at
a different~$z_{{m_Q}_i}$, then one has a system with different quark
masses~$m_{Q_i}$.

\bigskip

\begin{figure}  
\begin{center}  
  \includegraphics[width=0.6\textwidth]{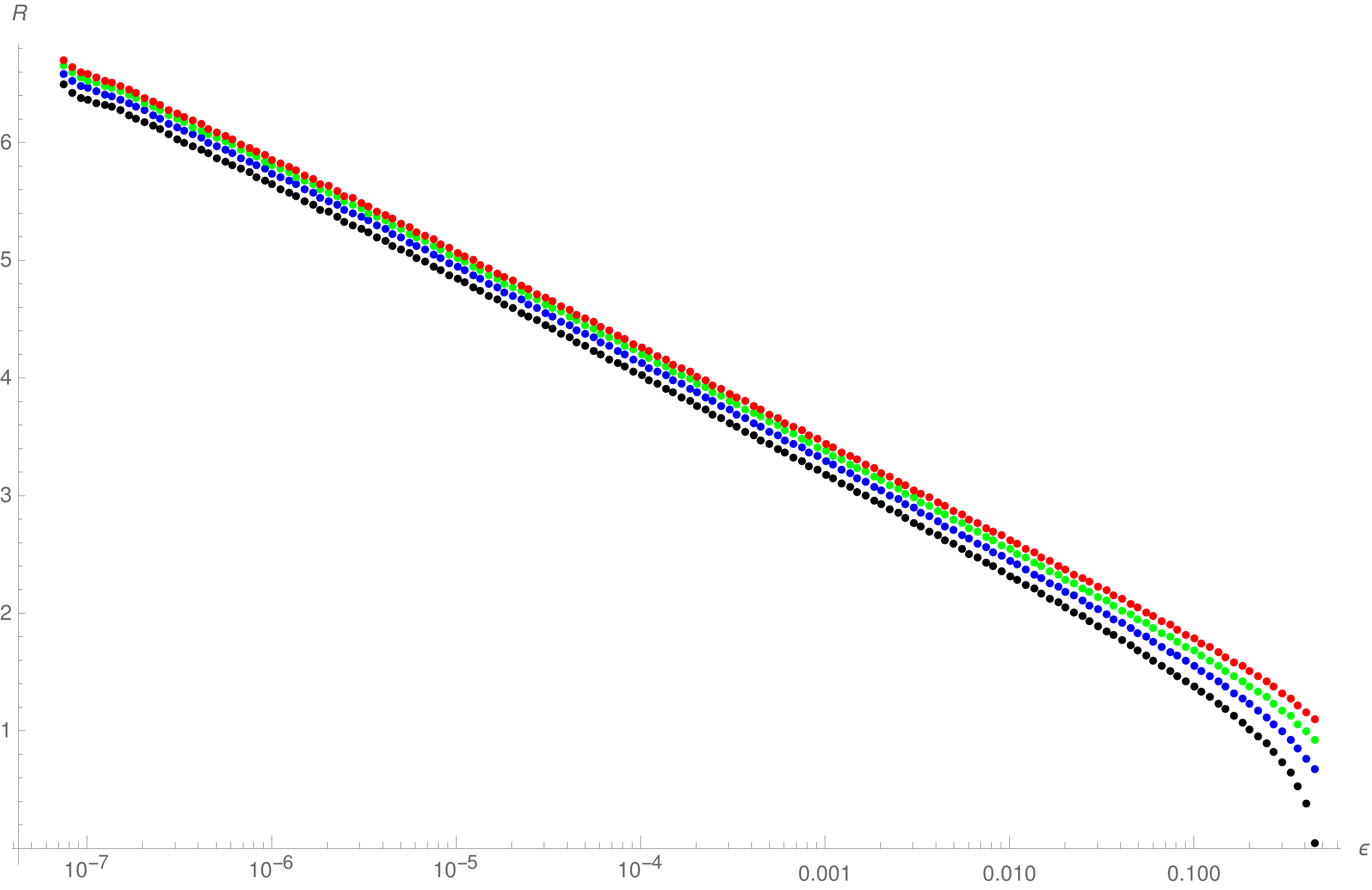}
  \caption{\label{outervsepsilon} Distance between the quarks (string
    endpoints) for a single loop $({\bf I})$, as a function of
    dimensionless separation of the tip of the loop from the wall
    $\epsilon=(z_{B}-z_\Lambda)/z_{\Lambda}$. Plots are given for
    different values of the quark massess: from bottom to top, black
    $m=1.31$, blue $m=2.03$, green $m=3.55$ and red $m=9.5$.}
\end{center}
\end{figure}

\begin{figure}  
  \begin{center}
    \vspace{2ex}
  \includegraphics[width=0.9\textwidth]{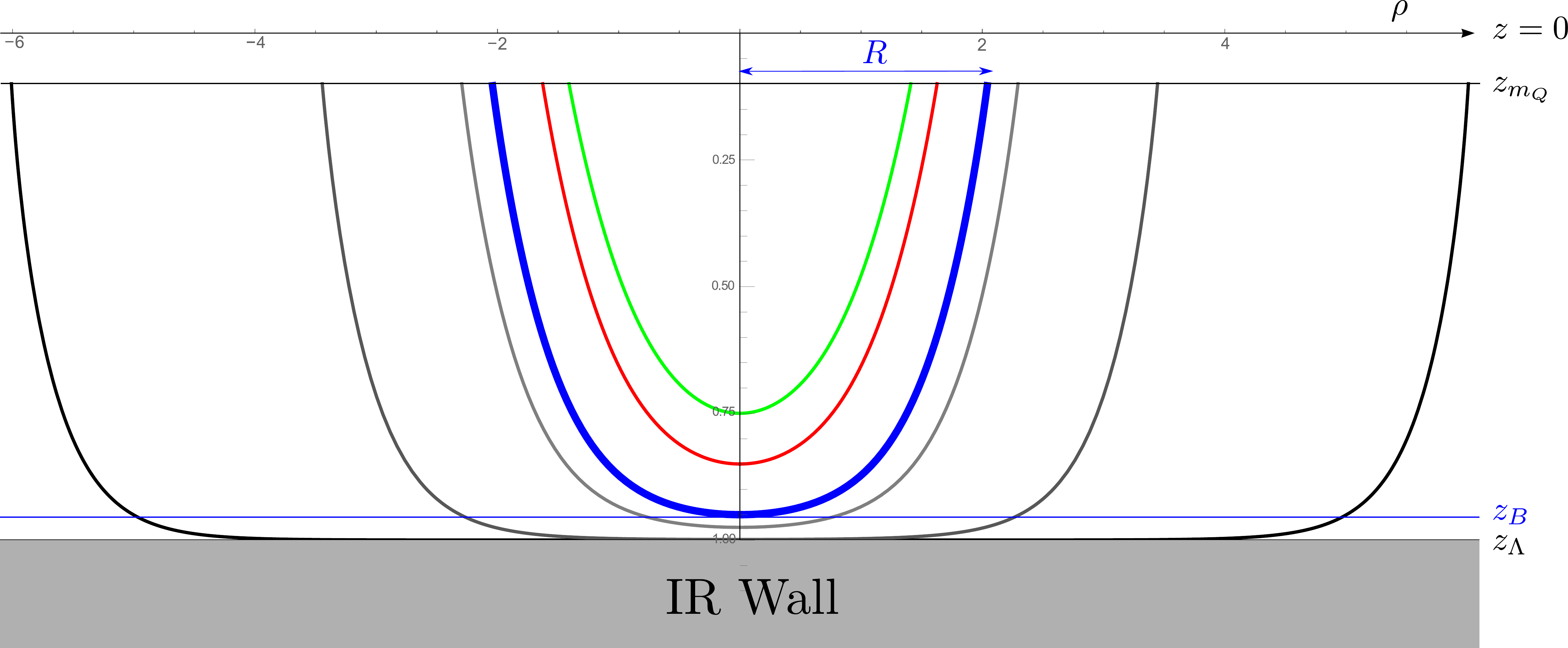}
  \caption{\label{loopssingle} One-parameter family of solutions of
    type ({\bf I}) plotted for a mass $m_Q=9.57$ of the external
    quarks.  Different solutions are parametrised by different
    distances between the quarks~$R$, or equivalently, different
    distances from the bottom of the loop to the wall. For all the
    plots we have set~$z_{\Lambda}=1$.}
\end{center}
\end{figure}

Let us now first look at the solution of type ({\bf I}). The equation
of motion~\eqref{equationsofmotionV1} are second order differential
equations, and as such have two undetermined constants of integration.
In order to see which parameters characterise a solution, let us look
at the~$z=\sigma$ gauge, since this is the simplest and the results
are independent of the gauge. In this gauge~$\sigma$ takes values in
$(z_{m_{Q}},z_B)$ where~$z_B$ is the position of the bottom of the
string loop, see figure~\ref{loopssingle}. For the solution~({\bf I})
we require that the tip of the loop is at the coordinate
origin~$\rho(z_B)=0$.  Also, as we are interested only in smooth
loops, we will require that at the bottom ${\rm d} z /{\rm d}
\rho\big|_B=0$. For a given position of the probe brane $z_{m_Q}$,
these two requirements uniquely fix the solution~({\bf I}). We are
therefore only left with two parameters which specify the
solution~({\bf I}): $z_B$, the position of the bottom of the loop and
$z_{m_Q}$, specifying the position of the top of the loop. In what
follows we will usually work with fixed masses of the outer
quarks~$m_{Q}$, or equivalently we will fix~$z_{m_Q}$. If one shifts
the bottom of the string $z_{B}$, this will change the distance
between the string endpoints, i.e.~the distance between the outer
quarks $R$ on the probe~D8.  As the position of the bottom of the loop
comes closer to the wall ($z_B \rightarrow z_{\Lambda}$) the distance
between the quarks $R$ becomes larger and larger, see
figure~\ref{outervsepsilon}.

In this near-wall limit, the single string loop looks more and more
like a U-shaped string, see figure~\ref{loopssingle}. Only in this
limit $z_B \rightarrow z_{\Lambda}$ is the identification of the
``vertical'' parts of the loop with quark masses~\ref{quarkmasses}
fully justified~\cite{Kinar:1998vq}. Also, only for these kind of
U-shaped strings is the effective tension of the horizontal part of
the string identified with $\sim \Lambda_{QCD}$, as the position of
the wall specifies $\Lambda_{QCD}$ in this model. As the bottom of the
loop approaches the IR wall one discovers that the action scales more
and more quadratically with the size of the loop, which is the
expected behaviour for a single circular Wilson loop in a confining
theory.

\bigskip

For the double loop solution (${\bf II}$), one needs to introduce two
probe D8-branes. The outer boundary of the string worldsheet will end
on the brane with position~$z=z_{m_Q}$. The position of this brane
fixes the mass~$m_{Q}$ of the original (heavy) quark pair. Before the
split, the original flux tube stretches between these two outer
quarks. When the flux tube breaks, an inner boundary is formed on the
string worldsheet.  As argued earlier, the inner boundary of the loop
ends orthogonally on the second brane which has position~$z=z_{m_q}$,
and this position fixes the mass of the \emph{produced}
quark-antiquark pair,~$m_{q}$. When considering solutions of
type~({\bf II}), we will fix these two parameters $m_Q$ and $m_q$ as
they are the parameters which are given in the dual gauge theory.

\begin{figure}  
  \begin{center}  
    \includegraphics[width=0.9\textwidth]{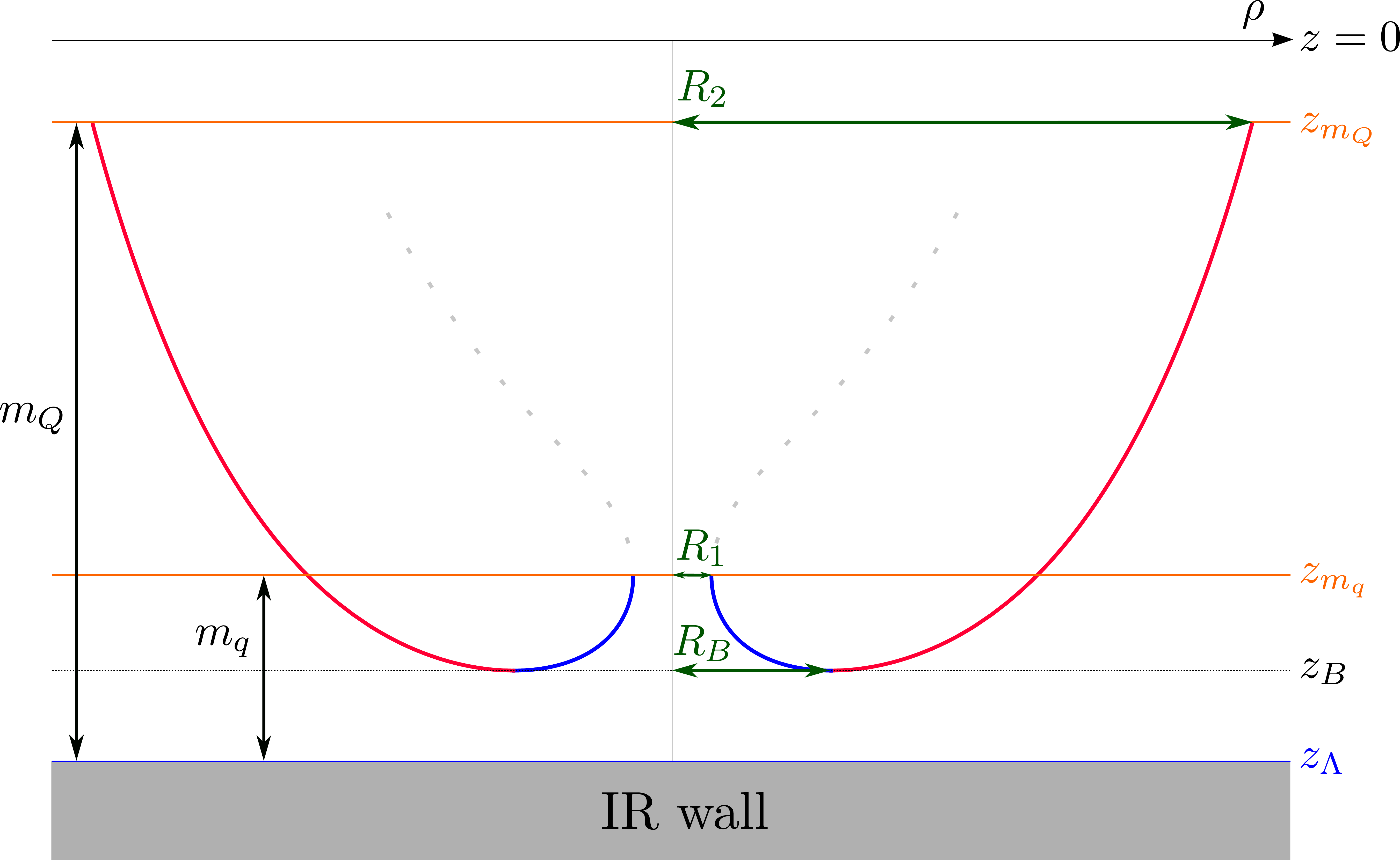}
    \caption{Slice of a generic double loop solution {(\bf II)} for
      $\phi=\text{const}$. The labels~$z_{m_Q}$ and~$z_{m_q}$ give the
      positions of two D8-branes and also specify the masses of the
      original (outer) quarks~$m_Q$ and pair-produced (inner)
      quarks~$m_q$. \label{looplabelled} }
  \end{center}
\end{figure}

Note that the solution (${\bf II}$) consists of two, outer and inner
branches, which are glued in a smooth way at the bottom of the loop,
$(z_B, R_B)$; these are coloured blue and red in  figure \ref{looplabelled}. In contrast to the loop
${\bf (I)}$, the bottom of the loop $({\bf II})$ is no longer at the
origin $\rho=0$, but it is placed at some point $( z_B, \rho = R_B
\neq 0)$.  Smoothness of the solution $({\bf II})$ at the bottom, as
before, implies $(d z / d \rho)|_{(z_B,R_B)} =0$. In principle, bottom
of the loop $z_B$ can be anywhere  between $0 \leq z_{m_Q} < z_{m_q}<z_B \leq z_{\Lambda}$.
However, we will be mainly interested in the
loops which have bottom near the wall $z_B \rightarrow z_{\Lambda}$,
since newly generated flux tubes are IR objects  which exist at
energies $\sim \Lambda_{QCD}$. It is also useful to introduce a
dimensionless parameter $\epsilon= (z_{\Lambda}- z_B)/z_{\Lambda}\ll1$.
%which tells us how far away from the QCD scale for a  particular
%solution.

Once the bottom of the loop (${\bf II}$) is fixed to some $z=z_B$, for
a given mass $m_q$, the inner (blue) branch of the solution is fully
fixed by the requirement of orthogonality of the string to the~$m_q$
probe brane (see~\eqref{orthogonal}) and the condition of smoothness
of the loop at the bottom. Therefore, the radius of the inner
loop~$R_1$ (on the~$m_q$ brane), as well as the radius~$R_B$ of the
bottom of the loop, are fixed once~$z_B$ and~$m_q$ are specified.

The outer (red) branch of the solution is fully fixed once the mass of
the original quarks~$m_Q$ is specified and one requires that this
branch is glued in a smooth way to the inner branch.  Note that the
outer branch of the solution~({\bf II}) need not end orthogonally on
the probe brane~$m_Q$, as the position of the outer quarks is fixed by
Dirichlet boundary conditions.  So in summary, solution~({\bf II}) is
fixed by specifying the quark masses $m_Q$, $m_q$ and the bottom of
the loop~$z_B$, or equivalently, $m_Q$, $m_q$ and the inner
radius~$R_1$ of the loop.

\begin{figure}  
  \begin{center}  
    \includegraphics[width=0.3\textwidth]{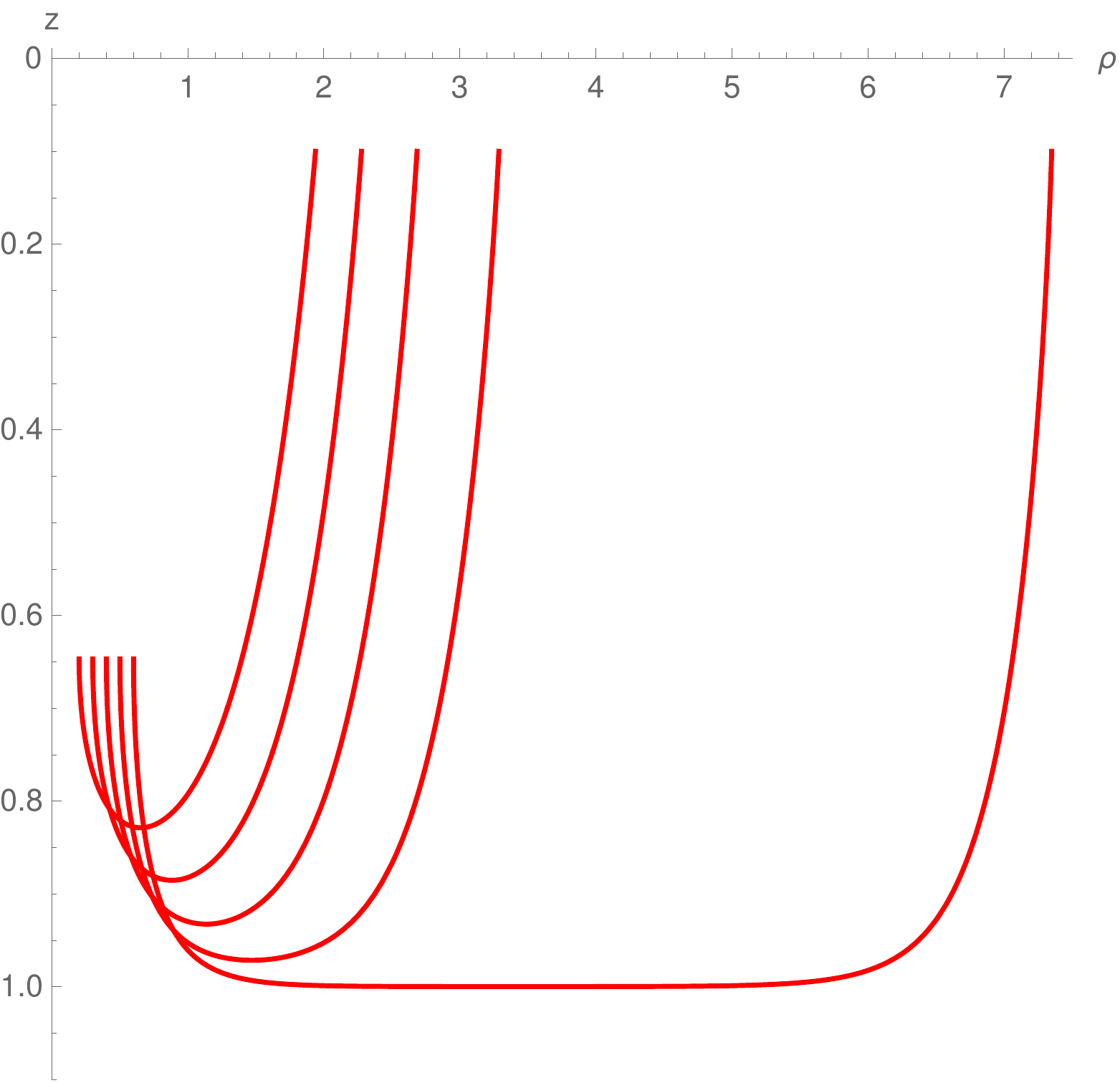}~
    \includegraphics[width=0.3\textwidth]{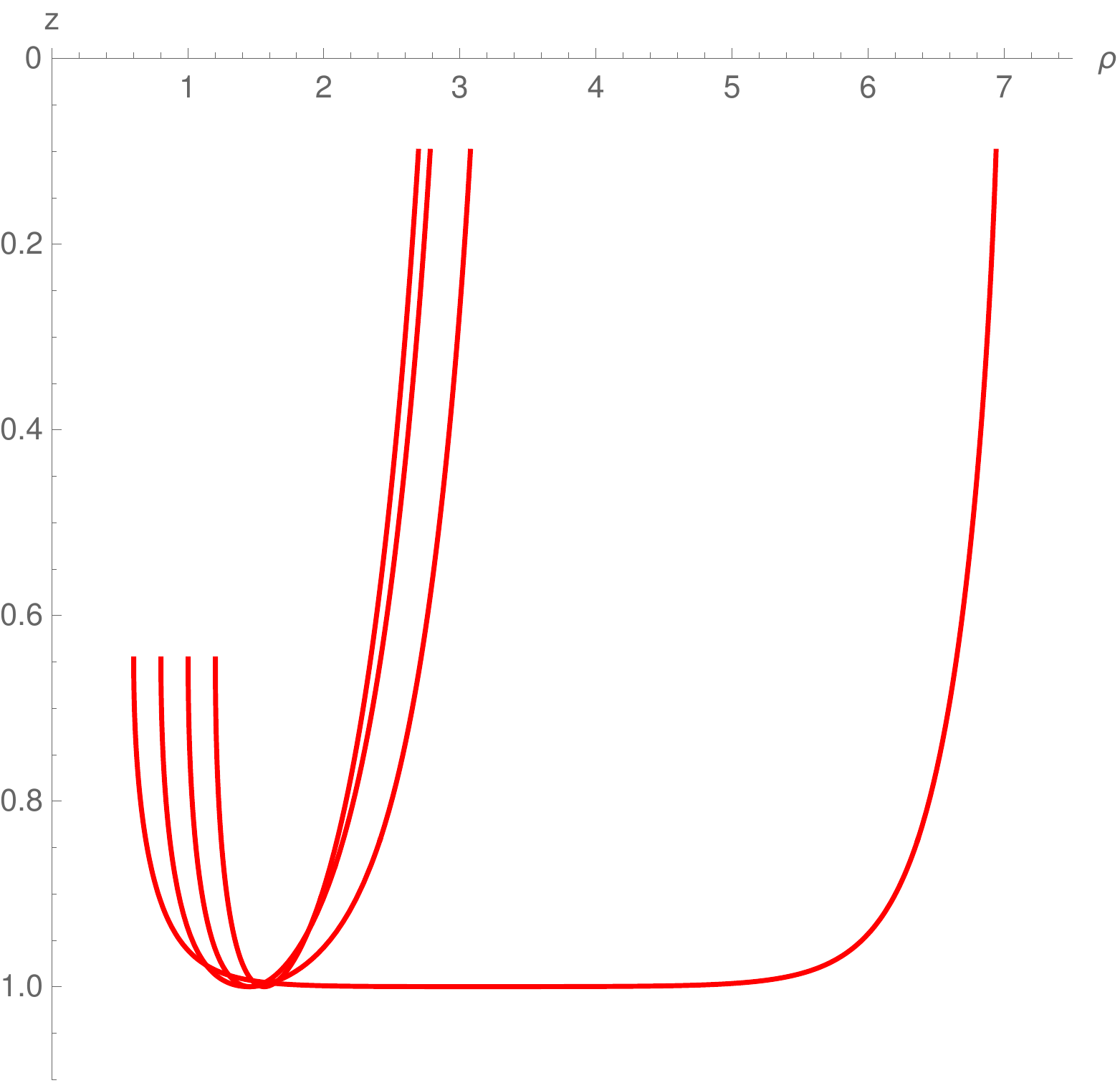}~
    \includegraphics[width=0.3\textwidth]{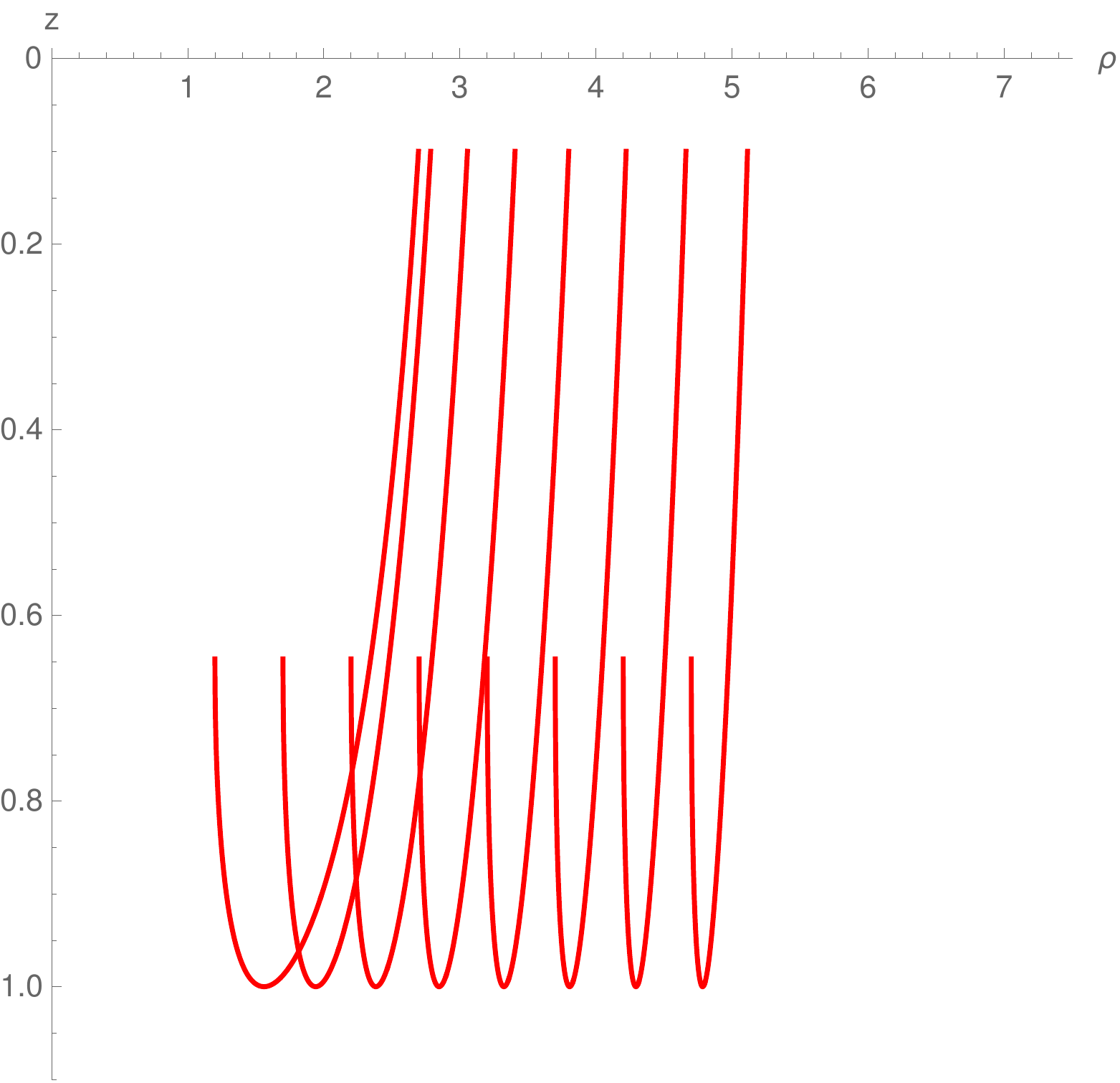}
  \end{center}
  \caption{Plots of the loops of type~({\bf II}) for increasing values
    of the inner radius~$R_1$. The left panel shows region~(i) until
    the value for which~$R_2$ is maximal. The middle panel shows loops
    in region~(ii), where $R_2$ decreases as $R_1$ increases,
    until~$R_2$ reaches its local minimum. The right panel shows the
    squashed loops of region~(iii) which occur beyond that.
    \label{limitlimit}
  }
\end{figure}

\begin{figure}
  \vspace{2em}
  \begin{center}  
    \includegraphics[width=0.45\textwidth]{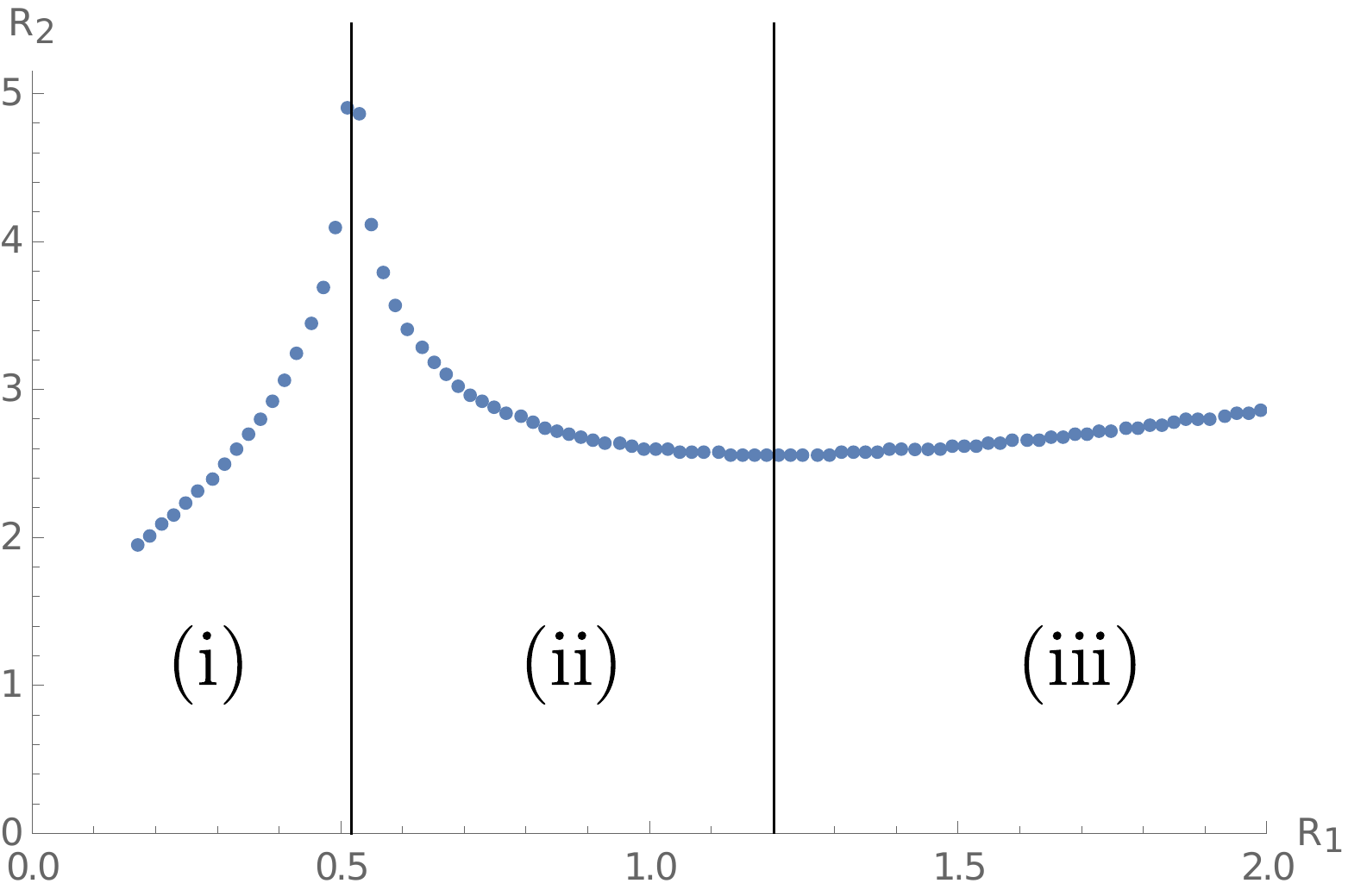}\quad
    \includegraphics[width=0.45\textwidth]{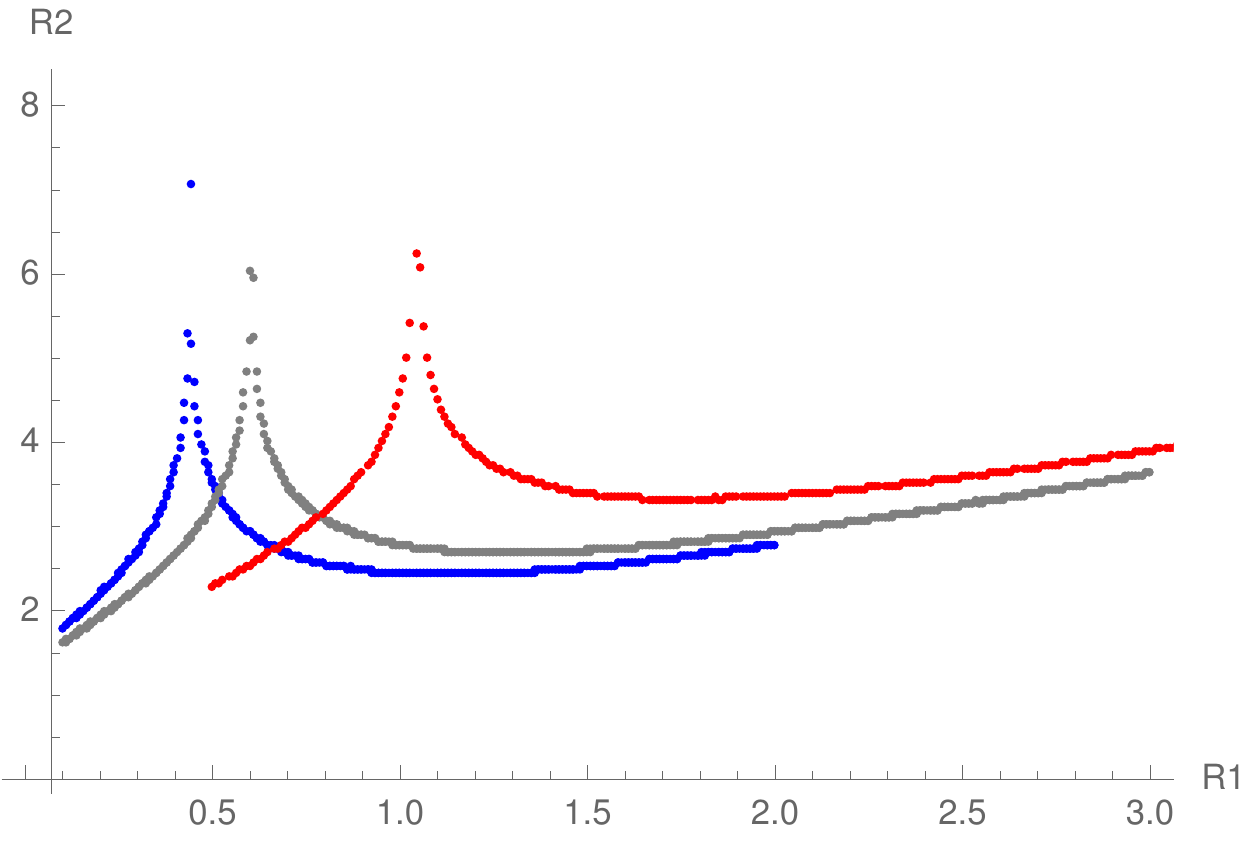}
  \end{center}
  \caption{The left panel shows the behaviour of the outer radius as
    inner radius varies, and the three different regions. Note that
    there are two critical points on this graph and the ``squashing''
    of loops takes place both in region (ii) and (iii). The right
    panel shows the behaviour for different
    masses~\mbox{$m_q=\{0.8,1.0,1.5\}$} (blue, gray, red) and fixed
    mass of the outer quarks~$m_Q=9.5$.  \label{varyingboth}}
\end{figure}

Let us now try to understand the moduli space of these double
loops. First we observe that for fixed~$m_q$ and $m_Q$, as the inner
radius $R_1$ is increased, the loop extends deeper and deeper into the
bulk, towards the IR wall, or in other words, $\epsilon$ decreases. As
this happens, \emph{initially} the loop becomes more and more U-shaped
and wider, with increasingly longer horizontal part and with larger
and larger outer radius~$R_2$. In this region, the effective size of
the system (the ratio of the outer versus the inner radius) grows. We
will refer to this region as region~(i). The left panel in
figure~\ref{limitlimit} shows a series of loops in this region. When
the inner radius~$R_1$ becomes larger than a particular critical
value~$R_{\text{crit-1}}$ the outer radius~$R_2$ starts to decrease as
the inner radius grows, so loops become more and more~\emph{squashed},
see the middle panel in figure~\ref{limitlimit}. Note that while
the squashing happens, the bottoms of all the squashed loops stay in
the region which is very close to the IR wall ($\epsilon \sim
10^{-5}$). We will refer to this region as region~(ii). Finally, when
the radius~$R_1$ becomes larger than another critical
value~$R_{\text{crit-2}}$, both inner and outer radius start to grow,
but the loop retains its squashed shape and it starts moving outwards
as a whole; see the right panel in figure~\ref{limitlimit}. We
will refer to this as region~(iii).  Figure~\ref{varyingboth} shows
the relation between the inner and outer radius as the inner radius
varies, in all three regions.  Note that as the inner quark mass is
increased the position of the peak moves to the right, but in a such a
way that the ratio of $R_{1}/R_{2}$ increases, so that the the system
is effectively at smaller volume. Put differently, systems with
smaller inner quark mass~$m_q$ have larger effective size in the sense
discussed above, and we expect them to reproduce the Schwinger results
more accurately. The peak in the~$R_2$ vs.~$R_1$ plot persists as~$m_q
\rightarrow m_Q$, but increasing the masses of outer and inner quarks
to larger value reduces the height of the peak and shifts its location
to larger values of $R_1$, so that eventually, for $m_q,m_Q
\rightarrow \infty$, only region~(i) remains and one is left with a
simple linear relation between $R_1$ and $R_2$, as expected from
e.g.~\cite{Armoni:2013qda}.

\begin{figure}
  \vspace{2em}
  \begin{center}
  \includegraphics[width=0.4\textwidth]{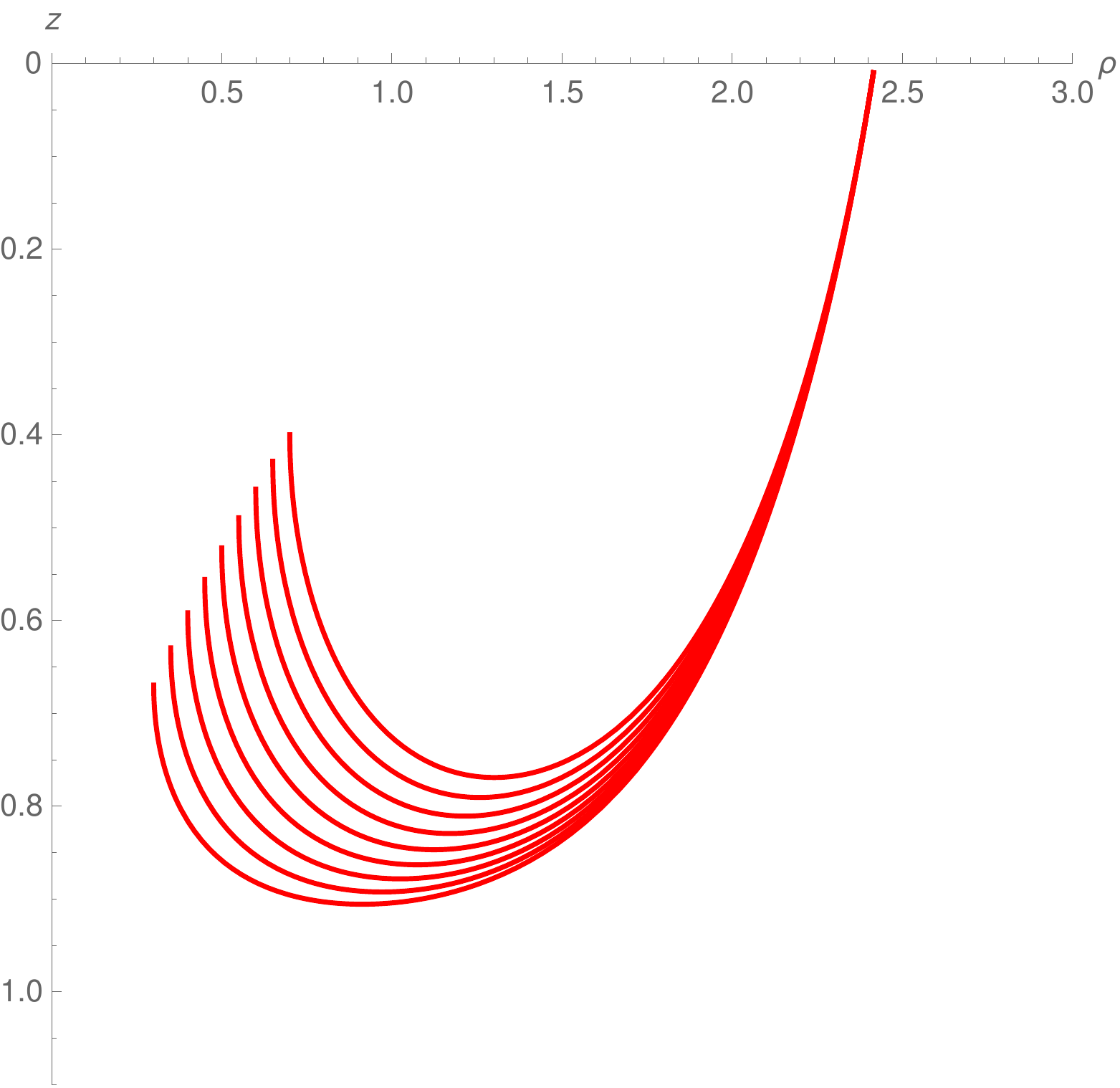}
  \caption{Shapes of double loops (in region~(i)) for fixed outer
    radius~$R_2=2.4$ and varying inner quark mass. Note that as the
    inner quark mass decreases, the radius at which these quarks are
    produced also decreases. \label{shapefixed} }
\end{center}
\end{figure}

\begin{figure}  
\begin{center}  
  \includegraphics[width=0.5\textwidth]{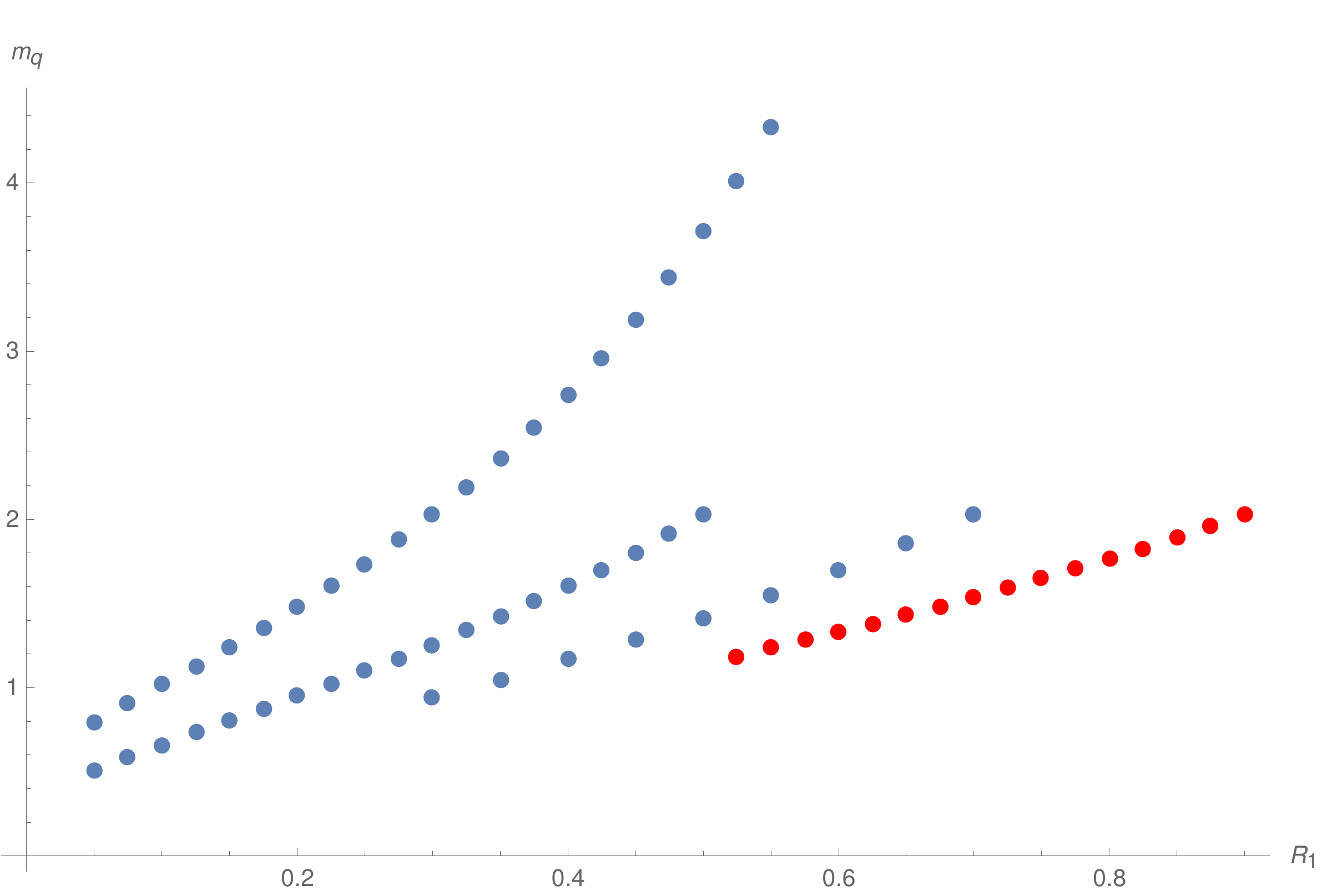}\caption{
    The relation between produced quark mass~$m_q$ and radius~$R_1$ at
    which the pair is produced for various loops in the region~(i). The
    plots are made for $m_Q=9.5$ and for different outer radii $R_2=
    \{1.6, 1.98, 2.37, 2.78\}$. The largest value of~$R_2$ corresponds
    to the bottom (red) curve.\label{shapefixed2}}
\end{center}
\end{figure}

In order to find the decay width of a given meson, we will need to
keep~$R_2$ and $m_Q$ fixed and look at the decay probability for
varying inner quark mass~$m_q$. Figure~\ref{shapefixed} illustrates
that, for strings in region~(i), the radius~$R_1$ at which the inner
quarks are produced decreases as~$m_q$ is
decreased. Figure~\ref{shapefixed2} shows the dependence between~$m_q$
and $R_1$ quantitatively, for different values of the outer quark
mass~$m_Q$. It shows that when the total system is smaller ($R_2$ is
smaller), quarks of the same mass~$m_q$ are produced at a radius~$R_1$
which is also smaller. Figure~\ref{shapefixed2} in addition suggests
that if $R_1\ll R_2$ the relation between $R_1$ and $m_q$ becomes linear,
as was the case for the Schwinger approximation (although the slopes
of these lines depends on the size of the system, unlike for
Schwinger). It is harder, and as we will argue below, less relevant,
to produce similar plots for regions~(ii) and~(iii), as for the
``squashed'' loops in these regions the dependence of~$R_1$ on~$m_q$
is rather weak (small variations of $m_q$ lead to almost no change in
$R_1$).

\subsection{Extracting the probability for decay}

Once the double loop solution is constructed we want to extract from
it the probability for the flux tube to break. As in the case of the
flat space instanton~\ref{actionproduction}, in order to compute this
probability, one first needs to compute the action of the solution
(${\bf II}$) and then subtract from it the action of the action of the
solution (${\bf I}$)
\begin{eqnarray}
  \label{actioninstanton}
  \Delta S(m_q,m_Q,R_2) = S_{II}(m_q,m_Q,R_2) - S_{I}(m_Q, R_2)\, \, . 
\end{eqnarray}
Both these actions are evaluated for the loops ${\bf (I)}$ and ${\bf
  (II)}$ which have identical outer radius~$R_2$, as this corresponds
to the physical ``size'' of the initial system
(cf.~footnote~\ref{f:nonsymmetric}). Generically, the probability for
a meson decay will depend on the size of the system~$R_2$ and on the
mass of the initial quarks~$m_Q$, in addition to the mass of the pair
produced quarks~$m_q$. The dependence of the decay rate on the
radius~$R_2$ is something that one expects for realistic mesons, as
the size of a meson is typically related to its angular momentum.

\begin{figure}  
\begin{center}  
  \includegraphics[width=0.5\textwidth]{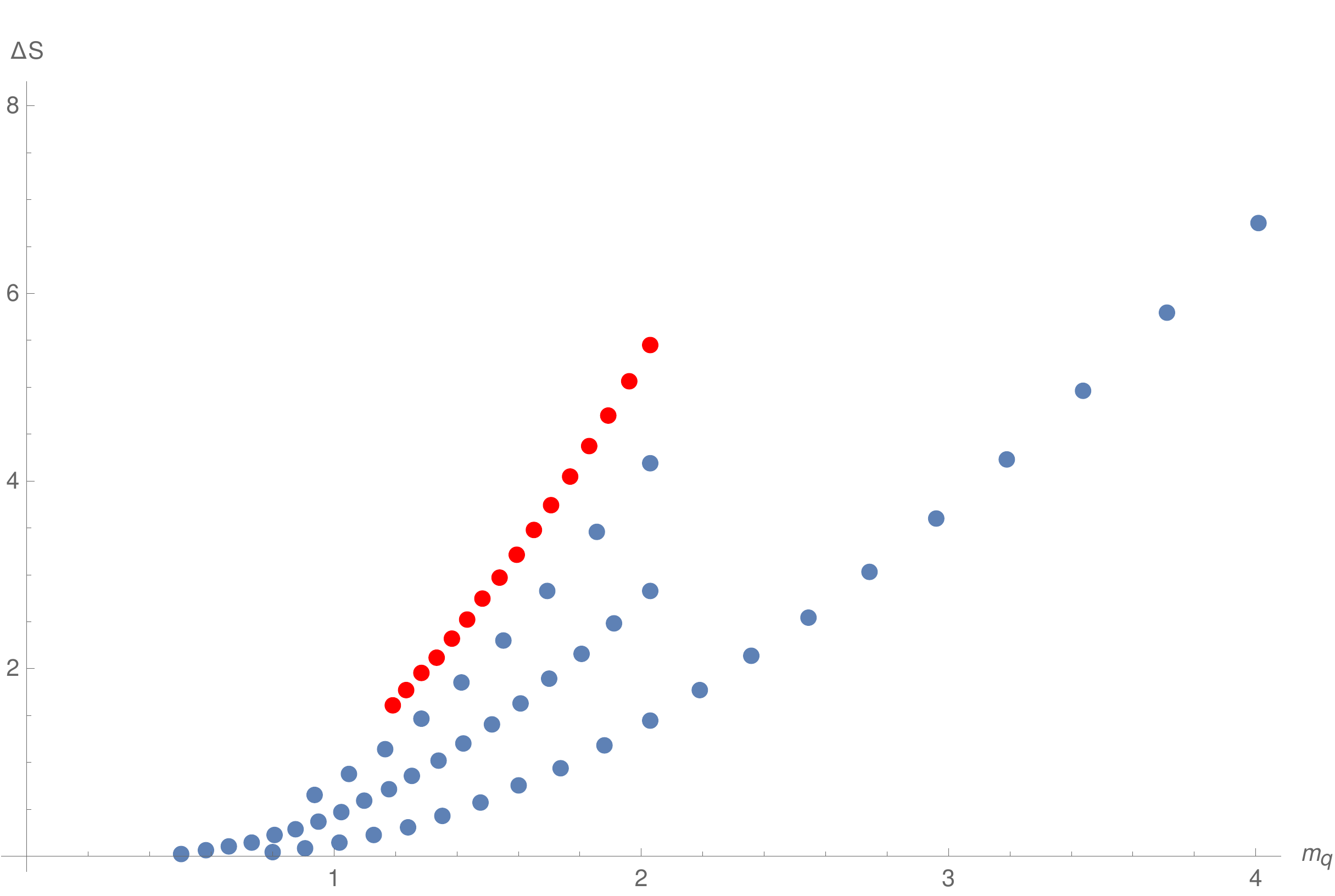}
\end{center}
\caption{\label{actionsvarious} Instanton action as a function of
  quark mass $m_q$, plotted for different size systems $R_2=\{1.6,
  1.98, 2.37, 2.78\}$. The upper curve (red) with the largest action
  corresponds to the largest~$R_2$.}
\end{figure}

As noted before, for a given system of fixed and large enough~$R_2$,
generically there are three possible radii at which quarks can be pair
produced, see figure~\ref{varyingboth}.  Each possible radius belongs
to one of the regions (i), (ii) or (iii).  Let us first analyse
instantons in the region~(i).  Figure~\ref{actionsvarious} shows the
instanton action~\ref{actioninstanton} in the region~(i) as a function
of quark mass. As the quark mass $m_q \rightarrow 0$, the instanton
action goes to zero, i.e.~lighter quarks are more likely to be
produced, as expected. We also see that the probability per unit time
and volume depends on the ``size'' of the system, and that the
probability for a larger meson to split is smaller than for smaller
mesons. This may sounds unintuitive, as one may expect larger
mesons to be more unstable.  However, one should keep in mind
that, when computing the lifetime of mesons, one still needs to
multiply this probability with the meson volume.

We should also note that when evaluating the action for instantons
using double loops, one always needs to make sure that the instanton
action is smaller than the action of two \emph{disconnected} loops
which have the same radii~$R_1$
and~$R_2$~\cite{Olesen:2000ji,Gross:1998gk}.  For all the loops
discussed in this paper, we have always checked that this holds so that
no Gross-Ooguri-Olesen-Zarembo type phase transition takes place.

\begin{figure}  
  \begin{center}  
    \includegraphics[width=0.55\textwidth]{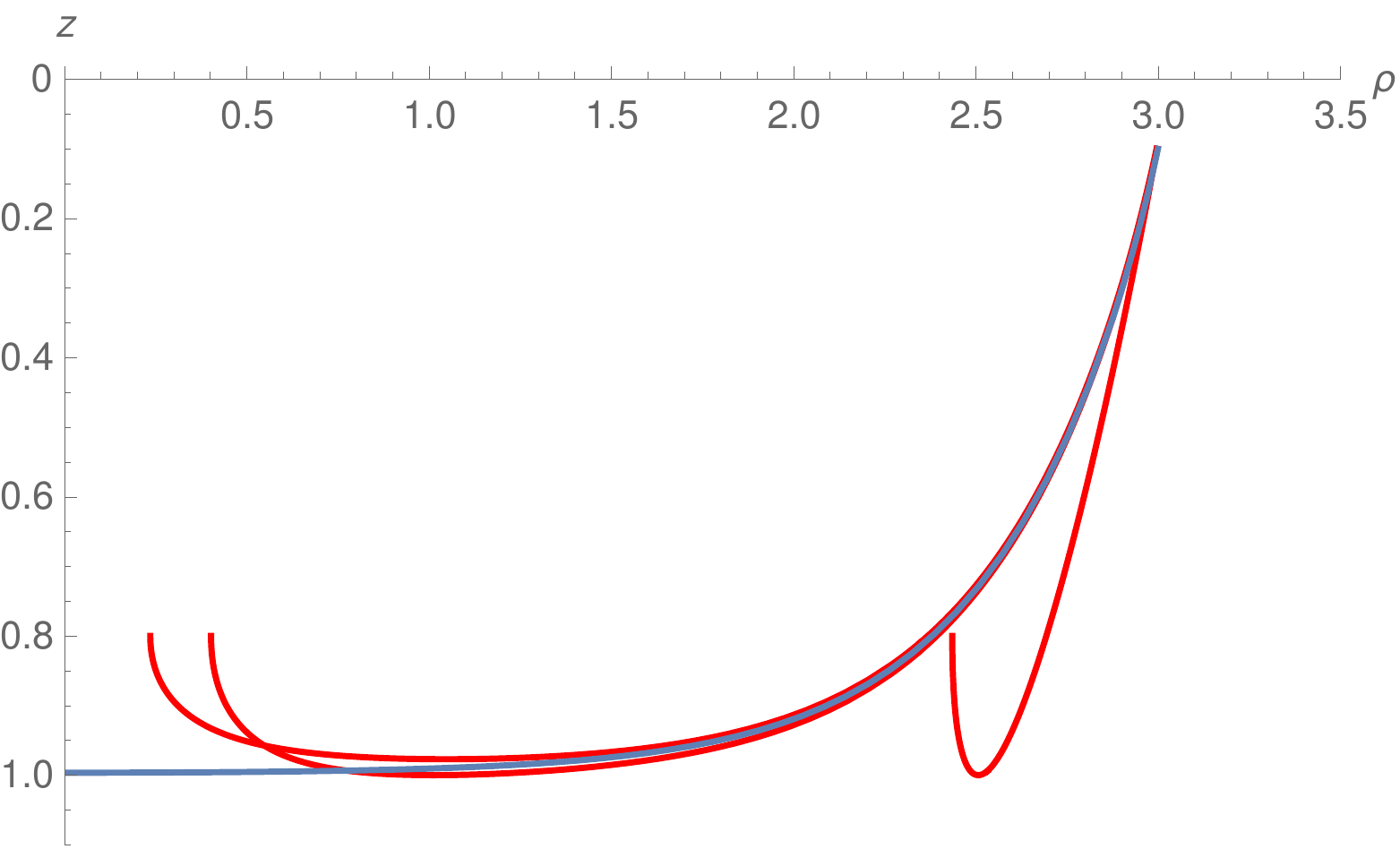}
  \end{center}
  \caption{\label{loopsregions} Double loops in the regions (i), (ii)
    and (iii) for the same outer radius~$R_2=3.0$ and inner quark
    mass~$m_q=0.62$ (corresponding to $z_{m_q}=0.8$). The blue curve
    extending all the way to~$\rho=0$ is a single loop with the same
    radius~$R_2=3.0$. }
\end{figure}

Evaluating the instanton action for radius $R_1$ in regions (ii) and
(iii) one gets that $S(R_{1\text{(i)}},R_2)
<S(R_{1\text{(ii)}},R_2)<S(R_{\text{1(iii)}},R_2)$.  We note however,
that the difference between these three actions is minimal, less than
a percent.  Hence it seems that decay in the region~(i) is the most
dominant, although only marginally.  Shapes of generic loops in
regions~(i), (ii) and (iii) which have the same $R_2$ are plotted in
figure \ref{loopsregions}.\footnote{At first glance it may look
  strange that loops in the region (ii) and (iii) have larger action
  than the action of the loop (i), given that loops (ii) and (iii)
  look like squashed version of loop (i). However, none of these loops
  is strictly rectangular and for given fixed $R_2$ they do not have
  tips which at the same distance from the wall.  So in order to
  compare actions one needs to evaluate them explicitely.}  It is
unclear to us at present what is the physical relevance of the
squashed loops, in particular the very squashed loops in
region~(iii).  These loops seems to suggest the existence of
``exotic'' decay channels for meson decay, where the pair-produced
quarks remove most of the flux tube in the decay. It could be that,
once the treatment of the angular momentum is taken properly into
account, these decays are forbidden due to selection rules. 

So in summary, the computation outlined above produces the probability
for the decay of mesons of a particular ``size''. It is hard to
compare our findings, even at qualitative level, with experimental
data, as these are mostly not known for higher spin mesons.  However,
one expects that in a particular limit, the holographic computation
should reproduce to the computations of CNN and Schwinger which were
outlined in section~\ref{s:flat_breaking}. Both these computations
work with a constant (chromo)electric field in infinite volume and in
the approximation where the produced quarks do not back-react on the
field.

\begin{figure}
  \vspace{4ex}
  \begin{center}  
    \includegraphics[width=0.5\textwidth]{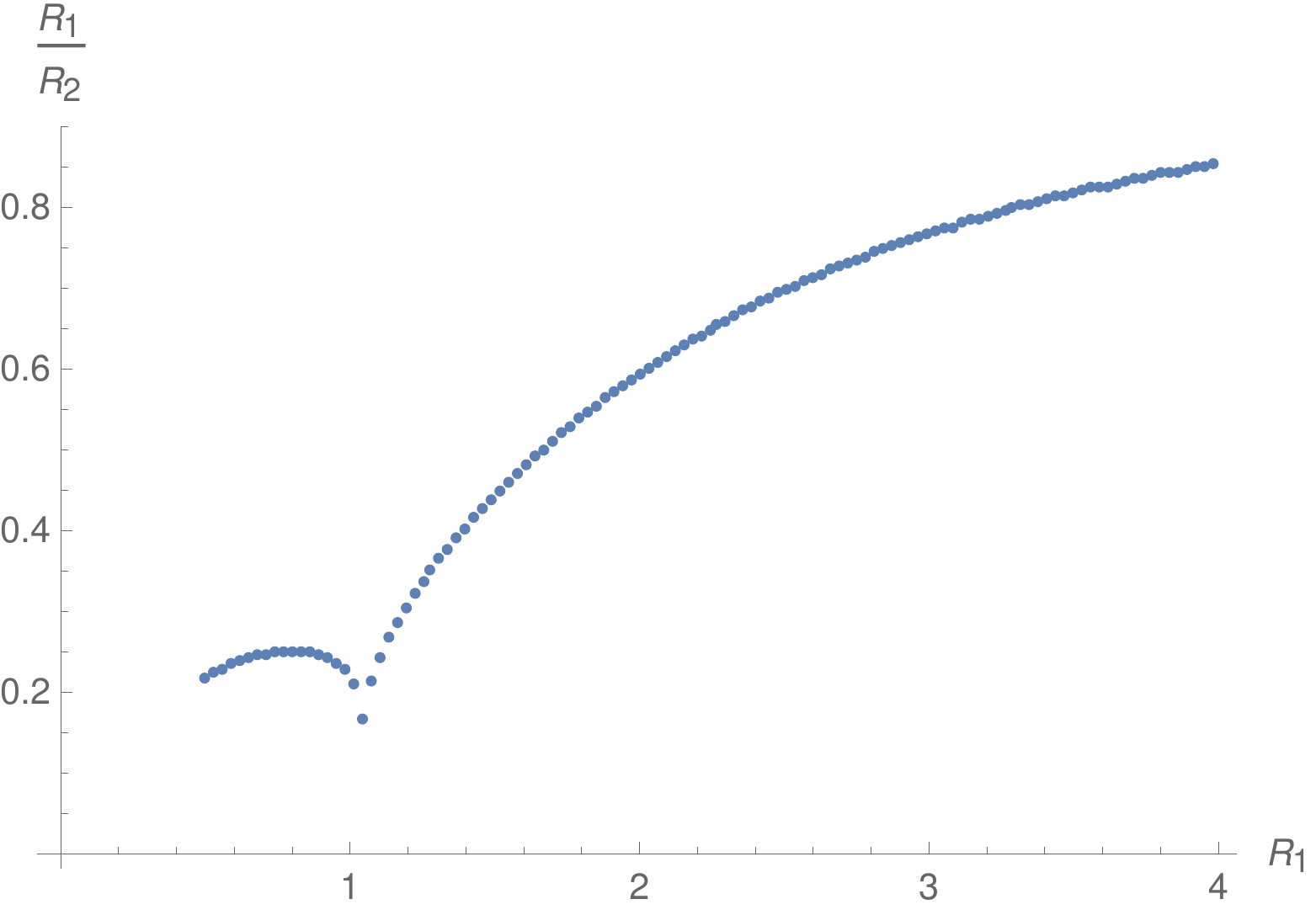}
    \vspace{-2ex}
  \end{center}
  \caption{\label{largevolume} Relative radius $R_1/R_2$ of the system
    as a function of the inner radius~$R_1$ for a fixed chosen
    mass~$m_1=1.5$. The largest effective volume is achieved for
    configurations near the boundary between region~(i) and~(ii) in
    the notation of figure~\ref{varyingboth}.}
\end{figure}

\begin{figure}[b!]  
\begin{center}  
  \includegraphics[width=0.9\textwidth]{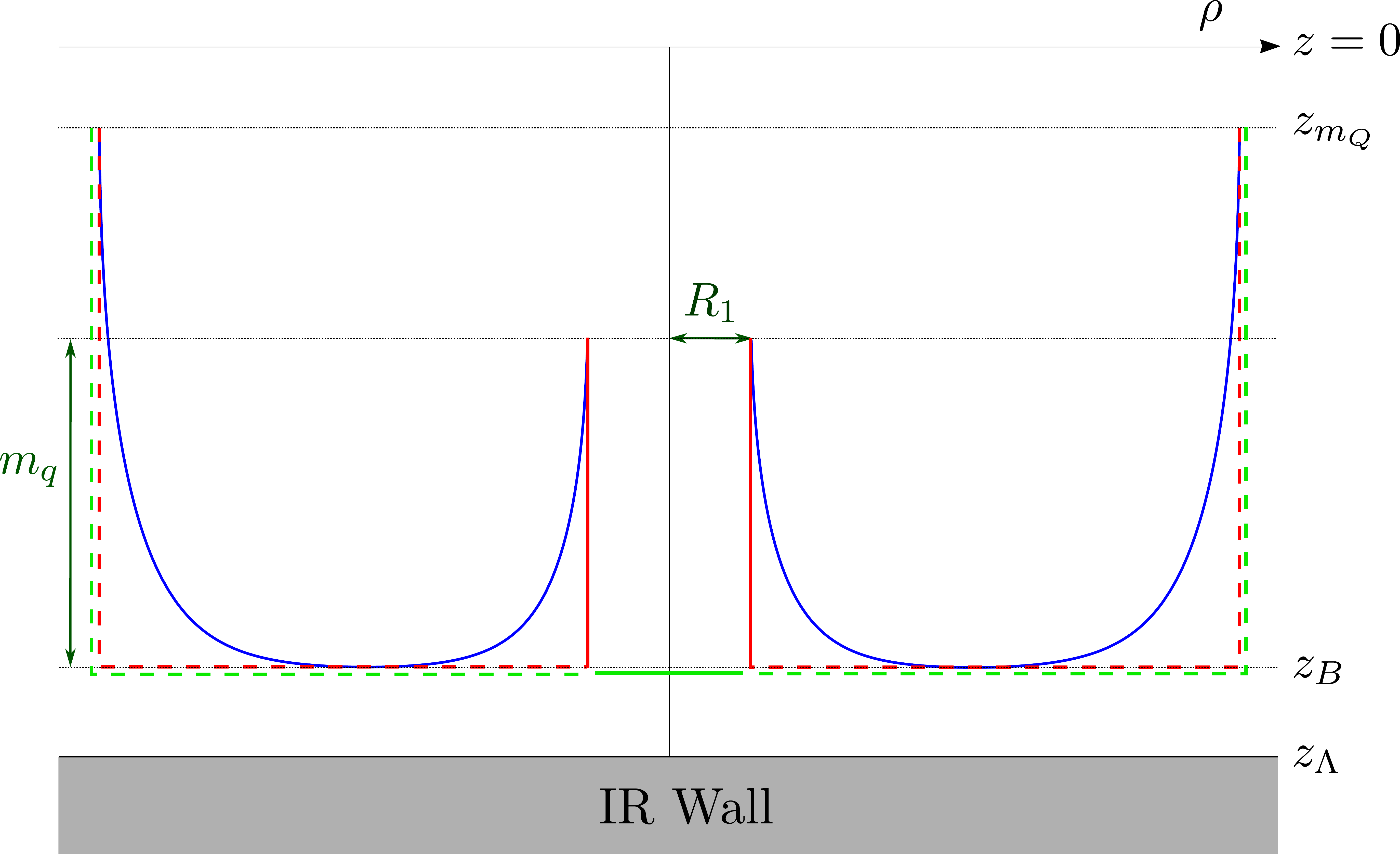}
  \caption{U-shaped loop approximation (red rectangular straight lines
    for the double loop, green straight lines for the single loop
    which is to be subtracted). The dashed parts are the same for the
    single and double loop solutions. For comparison we have also
    displayed an actual solution (blue curve). \label{loopapproximate}
  }
\end{center}
\end{figure}

In order to achieve a large-volume limit in the holographic set up,
one needs to look at long strings, with a horizontal part which is as
large as possible in comparison to the
radius~$R_1$. Figure~\ref{largevolume} shows the effective size of the
system, i.e.~the distance~$R_1$ at which the quarks are produced
versus the full size of the system~$R_2$ for quarks of fixed
mass~$m_q$.  We see that the largest effective volume is achieved for
loops at the boundary between the regions~(i) and~(ii). Note that
having the largest effective volume does not mean that the instanton
action for these loops is the smallest for a given~$m_q$ fixed.  By
looking at shapes of these large-volume loops, see
figure~\ref{loopsregions}, we see that they look most stretched and
moreover they look most like rectangular U-shaped loops. In order to
get an idea of what we should expect the action for these loops to
look like in the full numerical solution, let us consider an
approximation of these large-volume loops with rectangular U-shaped
loops, as indicated in figure~\ref{loopapproximate}.  In this
approximation, the action of the outer part of the loop (the dashed
red segments in the figure) is the same for single and double loops
and does not contribute to the instanton action. We therefore find that
\begin{equation}
  \Delta S_{\text{inst}} \sim S_{\text{tube}} - S_{\text{disc}} \sim 2 \pi R_1 m_q -
  R_1^2 \pi T\,,
\end{equation}
where we have used that $m_q$ is proportional to the height of the
tube, see~\eqref{quarkmasses}. This expression is similar to the one
obtained in the world-line derivation~\eqref{e:worldlineaction}, and
similar to the flat space expression~\eqref{actionproduction}, except
that in those expressions the mass~$m_q$ and the radius~$R_1$ are
linearly related through the equations of motion. In the present case,
however, the mass~$m_q$ and the radius~$R_1$ are \emph{independent} since
the outer radius is (by construction) decoupled from the rest and is
hence arbitrary. Our crude approximation therefore almost, but not
completely, reproduces the Schwinger approximation.

\begin{figure}  
  \begin{center}  
    \includegraphics[width=0.5\textwidth]{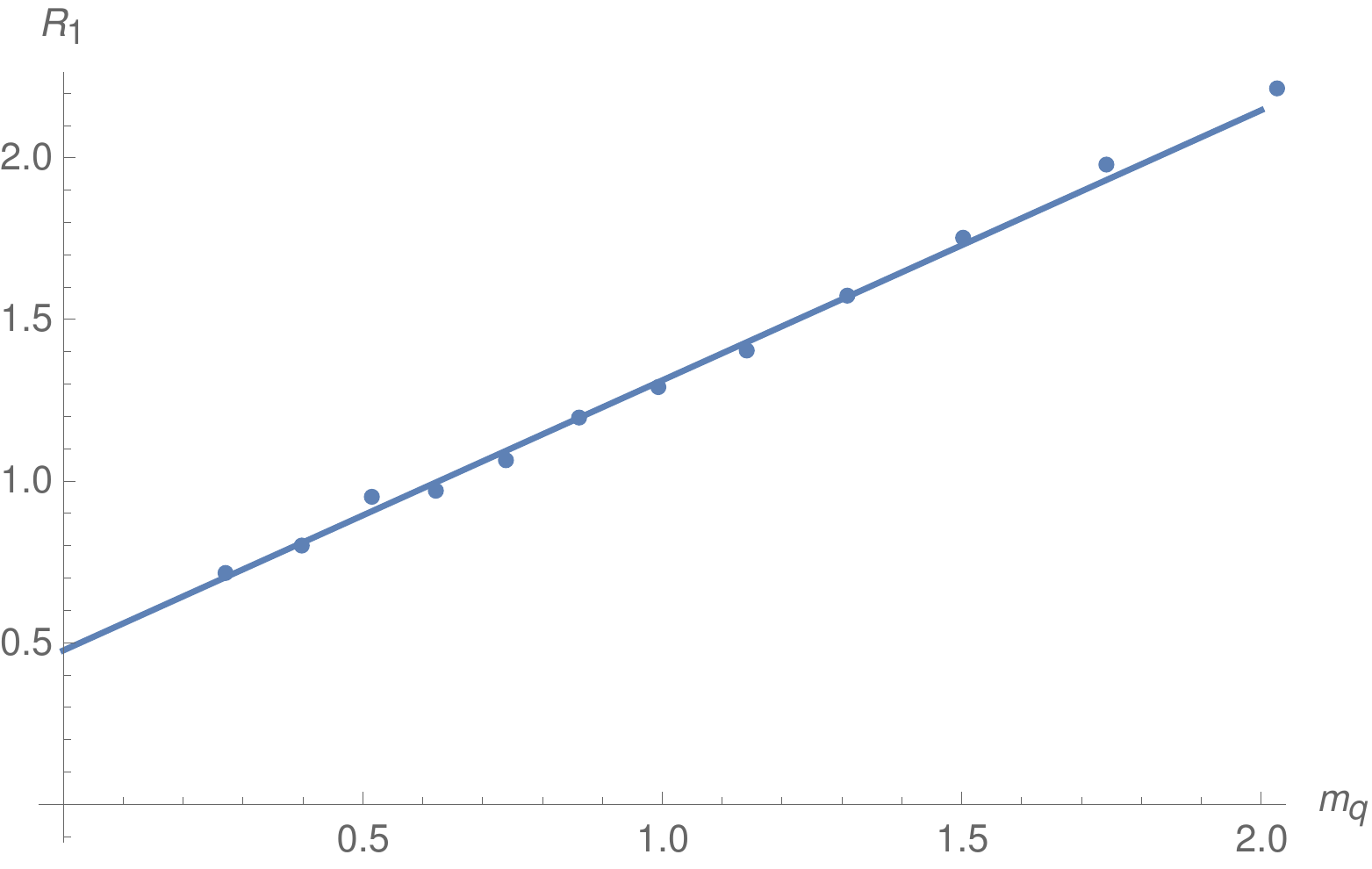}
  \end{center}
  \caption{Linear relation between the radius at which quarks are
    produced and their mass, obtained from the series of ``maximal''
    loops on the boundary between regions~(i)
    and~(ii). \label{alsovarious}}
\end{figure}

Motivated by this discussion, we will now focus attention to strings
with the largest possible volume (largest ratio of~$R_1/R_2$), which
corresponds to the peak between region~(i) and~(ii) of
figure~\ref{limitlimit}, for small quark mass~$m_q$. For these strings
we find a nice linear relation between~$R_1$ and~$m_q$, see
figure~\ref{alsovarious}. Note, however, these strings do not all have
the same outer radius~$R_2$. In order to nevertheless compare their
actions, we will need to focus only on the inner part of the loop
(from~$\rho=R_1$ to $\rho=R_B$). The outer parts of the single and
double loop (from~$\rho=R_B$ to $\rho=R_2$) are approximately equal
(as they are in the caricature) and would therefore cancel in the
instanton action after subtracting the single loop background (see
the loop in region~(i) in figure~\ref{loopsregions}).  As in the caricature, we will furthermore
approximate the shape of the single loop between $\rho=0$ and
$\rho=R_B$ by a straight line segment at constant $z=z_B$.

\begin{figure}  
  \begin{center}  
    \includegraphics[width=0.5\textwidth]{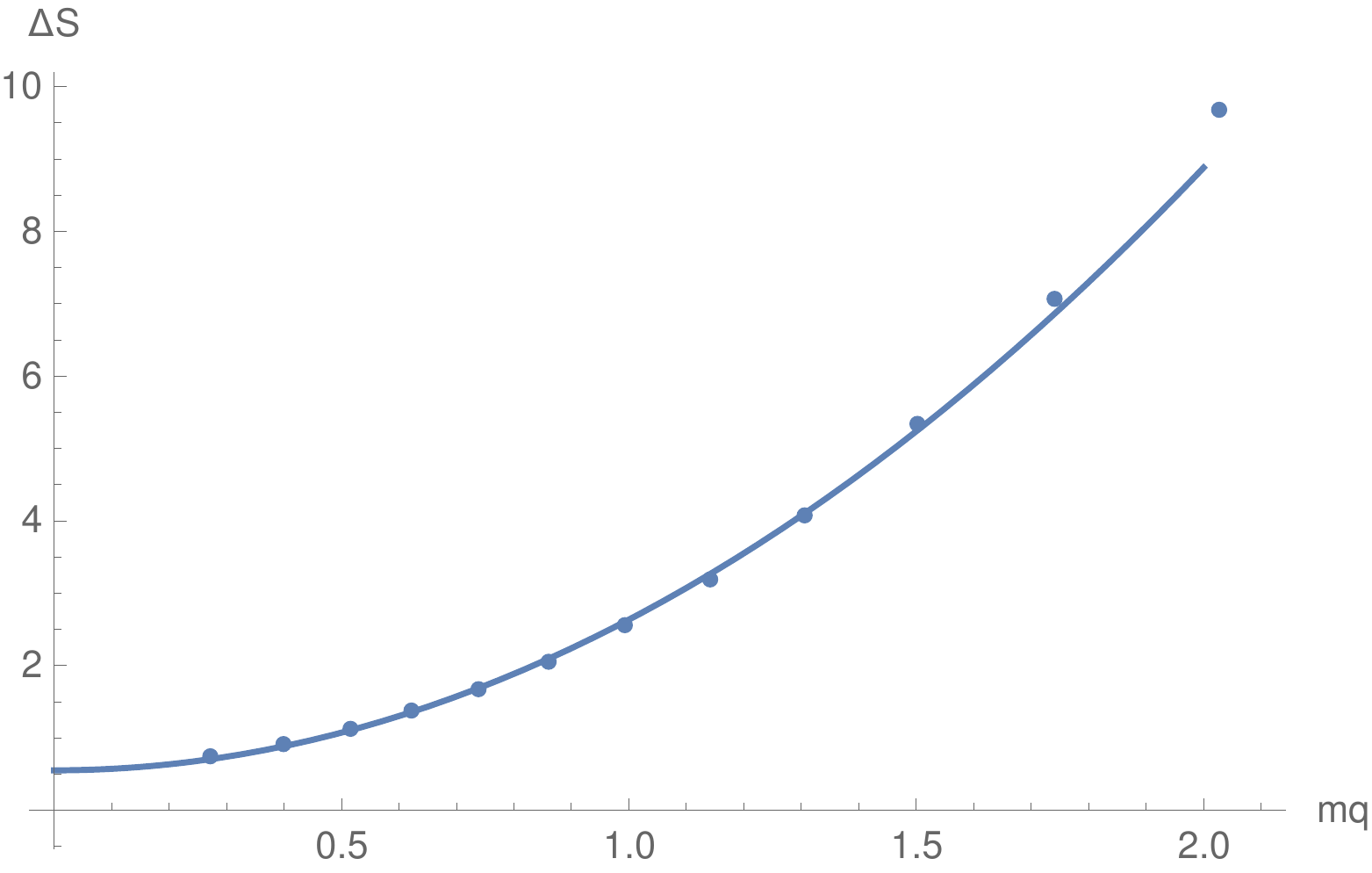}
  \end{center}
  \caption{\label{Fittinglines} Quadratic dependence of the instanton
    action~$\Delta S$ on the mass~$m_q$ of the pair-produced quarks in
    the `infinite volume limit approximation' in which one considers
    only loops near the boundary between region~(i) and~(ii).}
\end{figure}

The above is an `infinite volume approximation' in the sense that it
holds when \mbox{$R_1 \ll R_2$}.  When we plot the instanton
action~$\Delta S$ versus the produced quark mass~$m_q$ for these
loops, one recovers a quadratic relation, see
figure~\ref{Fittinglines}.  While the best fit is quadratic, like for
Schwinger, there is a nonvanishing constant present. One could remove
such a term by modifying the normalisation, and it would anyhow be
cancelled in a computation of the ratio of any two probabilities, so
it is irrelevant.  We should also comment that the numerical factor in
front of~$m_q^2$ in this fit differs from the factor of $\pi/4$ in the
Schwinger/CNN formula~\eqref{QCDSchwinger}, which is what one expects.
The expression~\eqref{QCDSchwinger} is valid only qualitatively in QCD
and also our holographic model is not a dual of real QCD, so one
should expect these kind of differences between the two results.

The probability~$P_{\text{pp}}$ for a flux tube to break, per unit length and
unit time, is obtained by exponentiation of the instanton action. In
the approximation of large mesons with an (infinitely long) flux tube,
translation invariance implies that the total probability for a meson
to split will be given by~$P_{\text{pp}}L$, where $L$ is the length of the
flux tube. In finite-size systems however, $P_{\text{pp}}$ will in general
depend on the position along the flux tube a well as the size of the
system. In order to evaluate the full probability we would first need
to construct the non-axially symmetric instantons, and then integrate
contributions of these instantons over the full length of the flux
tube. We leave such an investigation for another project.


\bigskip

\section{Discussion and outlook}

In this paper we have studied decay of mesons using instanton
techniques in two models: the old string model and the holographic
model of Sakai-Sugimoto. In the first model mesons are represented by
a pair of \emph{massive} particles connected with a relativistic
string (flux tube) in \emph{flat space}. In order to study their
decays, we have analytically constructed the worldsheet instanton
configuration which interpolates between two mesonic particles.  Using
this instanton, we were able to reproduce, up to an overall numerical
factor, the formula~\eqref{QCDSchwinger} for the probability of
breaking the QCD flux tube, derived a long time ago by Casher et
al.~\cite{Casher:1978wy}.  They derived their formula by making a
direct replacement of various quantities in the QED formula with the
analogue quantities in QCD. In contrast, in our approach the
connections between the theories followed, rather than being
postulated. After comparing the results for the decay probabilities it
followed that the chromoelectric field had the same role as an elastic
string. However this string does not couple in a minimal way to the
massive particles at the string endpoints, unlike the electric or
chromo-electric field in the worldline derivation of Schwinger-like
formula. Our derivation is very simple, yet it produces quite a
non-trivial field theory result. However, it was rudimentary in the
sense that we have restricted our attention to planar processes, where
both in- and outgoing particles lie in the same plane.  It would be
interesting to generalise this derivation to allow for the presence of
transverse momenta of the outgoing particles.

All of the flat space models (whether Schwinger, CNN or old string)
share one ``feature'': in order to incorporate (pair production of)
quarks one has to introduce an extra term in the action, by hand. In
the holographic model, in contrast, there is an unified treatment of
the flux tube and quarks. In the second part of the paper we have
developed a framework for studying the decay of large-spin mesons
using the holographic Sakai-Sugimoto model. We have constructed a
family of worldsheet instanton configurations, which interpolates
between incoming large-spin meson and two outgoing large-spin mesons.
The generic instanton describes the decay in which both the finite
size of the system and the backreaction of the pair-produced particles
is taken into account.  In this sense, the set up is more powerful
than either CNN or Schwinger computations which only give the
probability in the large volume limit and with no backreaction taken
into account. A shortcoming of our computation is that it is
restricted to (almost) cylindrically symmetric decay channels.  In the
infinite volume limit the probability is the same for breaking at all
points, but in a finite-size system we expect the probability to be
different along the flux tube. It would be very important to study
less symmetric decay.

Another restriction of our computation is that the outer (original)
quarks were considered only on circular orbits, which is a boundary
condition suitable for event generators where quarks accelerate away
from each other. In order to compute decays for mesonic particles in
general, one would need to generalise our computations to systems in
which the external quarks follow straight lines, and construct
instantons which are positioned at an arbitrary point along the flux
tube. Only than, total decay rates and the life times of the mesonic
particles can be computed and one could investigate interesting
prediction of the flat space, that $\Gamma/L$ is universal. 

When constructing instanton configurations for finite-size mesons, we
have discovered an interesting decay channel in which the
pair-produced quarks ``eat'' most of the flux tube, leading to very
short outgoing mesons. At the moment it is not clear to us what is the
physical significance of such a decay channel. One possibility is that
once the angular momentum is properly taken into account (as it is not
in our Euclidean framework) these exotic decay channels will be
forbidden by a selection rule. It would be interesting to investigate
this question in the future. Related to this is the question of proper
treatment of angular momentum of mesons in the holographic setup. One
expects that angular momentum modifies decay rates, as it provides an
extra centrifugal potential for pair-produced quarks.  Some of these
effects have been investigated for the Schwinger process
in~\cite{Gupta:1994tx}.

In order to cross check our computation, we have also investigated
meson decays in the large-volume limit. As expected, we have
rediscovered the qualitative form of the Schwinger/CNN formula, up to
a numerical factor. The long strings which were used in this
large-volume limit offer a natural playground in which one could try
to set up a systematic study of finite-size effects. Namely, for
large, but finite-size systems, the probability for a string to decay
should have an expansion in powers of~$R_1/R_2$, where $R_1$ is the
radius at which quarks are produced and $R_2$ is the size of the
system. It would be interesting to quantitatively study this expansion
using the holographic set up.

It would also be interesting to extend our analysis to finite
temperature field theory. By introducing a horizon in the
Sakai-Sugimoto setup one could generalise our instanton configuration
to this background, and obtain the thermal probability for a flux tube
to split.  Finally, one could ask to which extend the results which we
obtain are dependent on the exact form of the metric. In particular,
one might ask if for instantons in other confining geometries the
quadratic dependence on the quark mass persists. We plan to
investigate these issues in future work.

\vfill\eject
%\bibliographystyle{kasper}
%\bibliography{kasbib}

\begin{thebibliography}{20}
\expandafter\ifx\csname natexlab\endcsname\relax\def\natexlab#1{#1}\fi

\bibitem[Sakai and Sugimoto(2006)]{Sakai:2005yt}
T.~Sakai and S.~Sugimoto, ``More on a holographic dual of {QCD}'', {\em Prog.\
  Theor.\ Phys.} {\bfseries 114} (2006) 1083--1118,
 \href{http://arxiv.org/abs/hep-th/0507073}{{\ttfamily hep-th/0507073}}.
%%CITATION = HEP-TH/0507073;%%.

\bibitem[{Sj\"ostrand}(1982)]{Sjostrand:1982fn}
T.~{Sj\"ostrand}, ``{The Lund Monte Carlo for jet fragmentation}'', {\em Comp.\
  Phys.\ Commun.} {\bfseries 27} (1982)
243.
%%CITATION = CPHCB,27,243;%%.

\bibitem[Andersson et~al.(1983)Andersson, Gustafson, Ingelman, and
  {Sj\"ostrand}]{Andersson:1983ia}
B.~Andersson, G.~Gustafson, G.~Ingelman, and T.~{Sj\"ostrand}, ``Parton
  fragmentation and string dynamics'', {\em Phys.\ Rep.} {\bfseries 97} (1983)
31.
%%CITATION = PRPLC,97,31;%%.

\bibitem[Casher et~al.(1979)Casher, Neuberger, and Nussinov]{Casher:1978wy}
A.~Casher, H.~Neuberger, and S.~Nussinov, ``Chromoelectric flux tube model of
  particle production'', {\em Phys.\ Rev.} {\bfseries D20} (1979)
179--188.
%%CITATION = PHRVA,D20,179;%%.

\bibitem[Armoni(2008)]{Armoni:2008jy}
A.~Armoni, ``{Beyond The Quenched (or Probe Brane) Approximation in Lattice (or
  Holographic) QCD}'', {\em Phys. Rev.} {\bfseries D78} (2008) 065017,
 \href{http://arxiv.org/abs/0805.1339}{{\ttfamily arXiv:0805.1339}}.
%%CITATION = ARXIV:0805.1339;%%.

\bibitem[Peeters et~al.(2006)Peeters, Sonnenschein, and
  Zamaklar]{Peeters:2005fq}
K.~Peeters, J.~Sonnenschein, and M.~Zamaklar, ``Holographic decays of
  large-spin mesons'', {\em JHEP\,} {\bfseries 02} (2006) 009,
 \href{http://arxiv.org/abs/hep-th/0511044}{{\ttfamily hep-th/0511044}}.
%%CITATION = HEP-TH 0511044;%%.

\bibitem[Sonnenschein and Weissman(2018)]{Sonnenschein:2017ylo}
J.~Sonnenschein and D.~Weissman, ``{The decay width of stringy hadrons}'', {\em
  Nucl. Phys.} {\bfseries B927} (2018) 368--454,
 \href{http://arxiv.org/abs/1705.10329}{{\ttfamily arXiv:1705.10329}}.
%%CITATION = ARXIV:1705.10329;%%.

\bibitem[Affleck et~al.(1982)Affleck, Alvarez, and Manton]{Affleck:1981bma}
I.~K. Affleck, O.~Alvarez, and N.~S. Manton, ``{Pair Production at Strong
  Coupling in Weak External Fields}'', {\em Nucl. Phys.} {\bfseries B197}
  (1982)
509--519.
%%CITATION = NUPHA,B197,509;%%.

\bibitem[Schwinger(1951)]{Schwinger:1951nm}
J.~S. Schwinger, ``On gauge invariance and vacuum polarization'', {\em Phys.\
  Rev.} {\bfseries 82} (1951)
664--679.
%%CITATION = PHRVA,82,664;%%.

\bibitem[Affleck and Manton(1982)]{Affleck:1981ag}
I.~K. Affleck and N.~S. Manton, ``{Monopole Pair Production in a Magnetic
  Field}'', {\em Nucl. Phys.} {\bfseries B194} (1982)
38--64.
%%CITATION = NUPHA,B194,38;%%.

\bibitem[Nayak(2005)]{Nayak:2005pf}
G.~C. Nayak, ``{Non-perturbative quark-antiquark production from a constant
  chromo-electric field via the Schwinger mechanism}'',
 \href{http://arxiv.org/abs/hep-ph/0510052}{{\ttfamily hep-ph/0510052}}.
%%CITATION = HEP-PH 0510052;%%.

\bibitem[Barbashov and Nesterenko(1990)]{Barbashov:1990ce}
B.~M. Barbashov and V.~V. Nesterenko, ``{Introduction to the relativistic
  string theory}'',
1990.
\newblock
%%CITATION = INSPIRE-304988;%%.

\bibitem[Semenoff and Zarembo(2011)]{Semenoff:2011ng}
G.~W. Semenoff and K.~Zarembo, ``{Holographic Schwinger Effect}'', {\em Phys.
  Rev. Lett.} {\bfseries 107} (2011) 171601,
 \href{http://arxiv.org/abs/1109.2920}{{\ttfamily arXiv:1109.2920}}.
%%CITATION = ARXIV:1109.2920;%%.

\bibitem[Bardeen et~al.(1976)Bardeen, Bars, Hanson, and Peccei]{Bardeen:1975gx}
W.~A. Bardeen, I.~Bars, A.~J. Hanson, and R.~D. Peccei, ``A study of the
  longitudinal kink modes of the string'', {\em Phys.\ Rev.} {\bfseries D13}
  (1976)
2364--2382.
%%CITATION = PHRVA,D13,2364;%%.

\bibitem[Kruczenski et~al.(2005)Kruczenski, Zayas, Sonnenschein, and
  Vaman]{Kruczenski:2004me}
M.~Kruczenski, L.~A.~P. Zayas, J.~Sonnenschein, and D.~Vaman, ``{Regge
  trajectories for mesons in the holographic dual of large-$N_c$ QCD}'', {\em
  JHEP\,} {\bfseries 06} (2005) 046,
 \href{http://arxiv.org/abs/hep-th/0410035}{{\ttfamily hep-th/0410035}}.
%%CITATION = HEP-TH 0410035;%%.

\bibitem[Kinar et~al.(2000)Kinar, Schreiber, and Sonnenschein]{Kinar:1998vq}
Y.~Kinar, E.~Schreiber, and J.~Sonnenschein, ``{$Q \bar{Q}$ potential from
  strings in curved spacetime -- classical results}'', {\em Nucl.\ Phys.}
  {\bfseries B566} (2000) 103--125,
 \href{http://arxiv.org/abs/hep-th/9811192}{{\ttfamily hep-th/9811192}}.
%%CITATION = HEP-TH 9811192;%%.

\bibitem[Armoni et~al.(2013)Armoni, Piai, and Teimouri]{Armoni:2013qda}
A.~Armoni, M.~Piai, and A.~Teimouri, ``{Correlators of Circular Wilson Loops
  from Holography}'', {\em Phys. Rev.} {\bfseries D88} (2013), no.~6, 066008,
 \href{http://arxiv.org/abs/1307.7773}{{\ttfamily arXiv:1307.7773}}.
%%CITATION = ARXIV:1307.7773;%%.

\bibitem[Olesen and Zarembo(2000)]{Olesen:2000ji}
P.~Olesen and K.~Zarembo, ``{Phase transition in Wilson loop correlator from
  AdS / CFT correspondence}'',
 \href{http://arxiv.org/abs/hep-th/0009210}{{\ttfamily arXiv:hep-th/0009210}}.
%%CITATION = HEP-TH/0009210;%%.

\bibitem[Gross and Ooguri(1998)]{Gross:1998gk}
D.~J. Gross and H.~Ooguri, ``{Aspects of large N gauge theory dynamics as seen
  by string theory}'', {\em Phys. Rev.} {\bfseries D58} (1998) 106002,
 \href{http://arxiv.org/abs/hep-th/9805129}{{\ttfamily arXiv:hep-th/9805129}}.
%%CITATION = HEP-TH/9805129;%%.

\bibitem[Gupta and Rosenzweig(1994)]{Gupta:1994tx}
K.~S. Gupta and C.~Rosenzweig, ``Semiclassical decay of excited string states
  on leading regge trajectories'', {\em Phys.\ Rev.} {\bfseries D50} (1994)
  3368--3376,
 \href{http://arxiv.org/abs/hep-ph/9402263}{{\ttfamily hep-ph/9402263}}.
%%CITATION = HEP-PH 9402263;%%.

\end{thebibliography}

\begingroup\raggedright\endgroup

\end{document}